\documentclass[12pt]{article}
\newlength{\abstractwidth}
\usepackage{mathrsfs,amsbsy,amssymb,latexsym,amsfonts,amsmath}
\usepackage[dvipdfm]{graphicx,color}

\newcommand\Lap{\bigtriangleup}
\newcommand\CH{\mathcal H}
\newcommand\vk{{\vec{k}}}
\newcommand\vx{{\vec{x}}}

\renewcommand{\thefootnote}{\fnsymbol{footnote}}
\renewcommand{\thanks}[1]{\footnote{#1}} 
\newcommand{\starttext}{
\setcounter{footnote}{0}
\renewcommand{\thefootnote}{\arabic{footnote}}}
\renewcommand{\theequation}{\thesection.\arabic{equation}}
\newcommand{\be}{\begin{equation}}
\newcommand{\bea}{\begin{eqnarray}}
\newcommand{\eea}{\end{eqnarray}}
\newcommand{\beq}{\begin{equation}}
\newcommand{\ee}{\end{equation}}
\newcommand{\eeq}{\end{equation}}

\catcode`\@=11
\@addtoreset{equation}{section}
\def\theequation{\arabic{section}.\arabic{equation}}
\catcode`@=12
\relax

\begin{document}
\renewcommand{\theequation}{\thesection.\arabic{equation}}
\begin{titlepage}
\bigskip
\rightline{KEK-TH-1502}
\rightline{KUNS-2368} 
\rightline{OIQP-11-10}

\bigskip\bigskip\bigskip\bigskip

\begin{center}
{\Large \bf Possible origin of CMB temperature fluctuations:\\}
{\Large \bf Vacuum fluctuations of Kaluza-Klein and string states\\
during inflationary era\\}
\end{center}

\bigskip\bigskip
\bigskip\bigskip

\centerline{Yoshinobu Habara$^{a}$, Hikaru Kawai$^{b}$, 
Masao Ninomiya$^a$, Yasuhiro Sekino$^c$}
\medskip
\medskip
\centerline{\small $^a$Okayama Institute for Quantum Physics, 1-9-1
Kyoyama, Okayama 700-0015, Japan}
\centerline{\small $^b$Department of Physics, Kyoto University, 
Kyoto 606-8502, Japan}
\centerline{\small $^c$Theory Center, Institute for Particle and 
Nuclear Studies, KEK, Tsukuba 305-0801, Japan}
\medskip
\medskip

\bigskip\bigskip
\begin{abstract}
We point out that the temperature fluctuations of cosmic microwave 
background (CMB) can be generated in a way that is different from
the one usually assumed in slow-roll inflation.
Our mechanism is based on vacuum fluctuations of fields which are 
at rest at the bottom of the potential, such as Kaluza-Klein modes
or string excited states. When there are a large number 
(typically of order $N\sim 10^{14}$) of fields with small mass in unit of
Hubble parameter during the inflationary era, this effect can give 
significant contributions to the CMB temperature fluctuations. 
This number $N$ makes it possible to enhance scalar
perturbation relative to tensor perturbation.  
Comparison with the observed amplitudes suggests that models with 
string scale of order $10^{-5}$ of 4D Planck scale are favorable.
\end{abstract}

\bigskip
\noindent
PACS numbers: 04.62.+v, 98.80.Cq, 11.25.-w, 11.25.Mj, 98.80.Qc

\bigskip
\bigskip

\noindent
{\small {\bf Email:} habara@yukawa.kyoto-u.ac.jp,
hkawai@gauge.scphys.kyoto-u.ac.jp,
ninomiya@yukawa.kyoto-u.ac.jp,
sekino@post.kek.jp}

\end{titlepage}
\starttext \baselineskip=18pt \setcounter{footnote}{0}


\section{Introduction}
Observation of cosmic microwave background (CMB) provides an
excellent opportunity for testing theories of high energy physics.
The CMB radiations are the photons emitted at the era of recombination
reaching us almost unscattered. It has a very homogeneous distribution
over the whole sky with thermal spectrum at $T\sim 2.7K$ with
fluctuations $\delta T/T$ of order $10^{-5}$. 
Temperature fluctuation is directly related to the gravitational 
potential at the last 
scattering surface by the relation, $\delta T/T=-\Phi/3$ 
(see e.g. \cite{Mukhanov}). 
Gravitational potential $\Phi$ is essentially frozen in the matter 
or radiation dominated universe, thus the observation of CMB enables 
us to trace back the universe to the era much earlier than 
recombination. 

It is believed that there has been a period of exponential expansion
(inflation) in the early universe~\cite{Guth}. Had the universe been decelerating
(matter or radiation dominated) since the beginning, the observable 
universe would have to be made of many spatial regions which have been 
initially independent, making it difficult to explain the 
homogeneity of our universe. 
Exponential expansion brings these regions in causal contact in the
past. This is the only compelling resolution of this horizon problem.

The fluctuations generated during inflation has nearly scale invariant
spectrum. At each time $\delta t\sim H^{-1}$ (where $H$ is the
Hubble parameter of inflation), fluctuations of order 
$\delta \phi\sim H$ will be created in a spatial region of 
horizon size $\sim H^{-1}$. This fluctuation is stretched by 
the cosmic expansion, and once the wavelength of fluctuation 
exits the horizon, it is frozen and treated classically. 
Quantum fluctuations continuously exit the horizon, and
this mechanism creates the same structure of perturbations
at every length scale (see e.g. \cite{Linde}). 

The observations of WMAP~\cite{WMAP} find a nearly scale 
invariant spectrum of primordial temperature fluctuations. 
It is often stated that WMAP confirmed inflation, and the 
results expected from PLANCK satellite will narrow down possible 
models of inflation. In making such a statement, it seems that a 
particular mechanism~\cite{slowrollfluctuation} 
for generating CMB fluctuations is assumed, 
which is based essentially on slow-roll inflation~\cite{slowroll}. 
Regarding the fact that theories of inflation have not been derived from
fundamental theory of quantum gravity yet, we believe it is
important to examine whether there are any issues 
that are overlooked
in making predictions from inflation.

In this paper, we point out that the temperature fluctuations 
of CMB can be generated by  purely quantum effects, which is
different from the mechanism usually assumed in the slow-roll scenario. 
Our mechanism is based on the vacuum fluctuations of a large
number of fields that are classically at rest at the bottom
of their potential. The effect from each field is small, but
we show that a sufficiently large number of fields from Kaluza-Klein (KK)
modes or string excitations can produce an observable level of
temperature fluctuations. The effect of these fields on the 
tensor perturbation is small. 

By comparing the temperature fluctuations obtained from
our mechanism to the observed amplitude, we find that 
a theory with relatively low fundamental scale 
(i.e. string scale being 5 orders of magnitude lower 
than the 4D Planck scale) is favored. 

\subsection{Temperature fluctuations}

To clarify the difference of our mechanism for generating
temperature fluctuations form the one in slow-roll scenario,
let us briefly review the latter~\cite{slowrollfluctuation}.

In most of the currently studied models of inflation, it is
assumed that a scalar field (inflaton) goes through a classical 
motion. In slow-roll inflation~\cite{slowroll}, inflaton rolls 
down the potential
which is flat enough for vacuum energy dominates over the
kinetic energy. In chaotic inflation~\cite{chaotic}, potential is generic
but the friction due to the expansion makes the motion
effectively slow. In other models such as  
N-flation~\cite{Nflation}, slowly moving classical field 
is effectively involved in certain sense. 

The equal value surface of the inflaton field provides a natural 
time slicing. Thus, fluctuations of inflaton $\delta\varphi$
can be reinterpreted as the fluctuation of time duration,
or how much the universe has expanded: The slice of
$\delta \varphi=0$ is obtained by gauge transformation,
$\delta t=- \delta\varphi/\dot{\varphi}_{\rm cl}$, and
on that slice, fluctuation of the spatial curvature 
${\cal R}=-H\delta t$ is generated\footnote{Although this 
heuristic derivation gives the correct answer in the slow-roll 
limit, for a consistent analysis, one should use the gauge 
invariant variable defined in \cite{MukhanovSasaki} which 
corresponds 
to the curvature perturbation on comoving hypersurfaces.
This enables one to study the general cases. 
See e.g.~\cite{LeachSasaki}.}
(which can be translated into the gravitational 
potential $\Phi$). 
In the slow-roll scenario, curvature perturbation is
enhanced due to the slowness of the classical motion 
$1/\dot{\varphi}_{\rm cl}$.
Using the fact that $\delta \varphi \sim H$, and the
slow-roll approximation, $\dot{\varphi}_{\rm cl}=V'/(3H)$,
curvature perturbation is written as ${\cal R}\sim 
(1/ \epsilon)(H/m_p)$,  
where $\epsilon=(V'm_p)^2/(8\pi V)^2$ is a slow-roll parameter
characterizing the flatness of the inflaton potential (see e.g.
\cite{LiddleLyth}).

During inflation, tensor perturbation is also generated~\cite{tensor}. 
In the linearized approximation around an isotropic background,
transverse-traceless tensor is decoupled from other fields.  
It satisfies massless equation of motion, and the amplitude
is of order $H/m_p$, as is clear from dimensional analysis,
where $m_p$ is the Planck scale. Tensor perturbation will produce 
B-mode polarization in the CMB. This is not observed
at present, and the tensor to scalar ratio is
bounded above by $r_{\rm T/S}=2 {\cal T}^2/{\cal R}^2
\lesssim 0.2$. This leads to an important conclusion that 
$H$ is at least 5 order of magnitude smaller than $m_{pl}$.

We propose a mechanism for generating temperature fluctuations
by vacuum fluctuations of the fields which are classically at rest. 
(Our mechanism is different from N-flation~\cite{Nflation} in 
this sense.)
Energy momentum tensor is quadratic in these fields, and their effect 
on the gravitational potential $\Phi$ is neglected in the usual first 
order perturbation theory. 
Each field gives a small contribution of order $(H/m_{pl})^2$ to $\Phi$, 
but when there are many fields (typically of order 
$N\sim 10^{14}$), 
this can sum up to an observable level.

The fields with small mass compared to $H$ do not 
oscillate during inflation, since the friction due to cosmic expansion
overdamps the oscillation. These fields contribute to temperature
fluctuations. When there are extra dimensions whose size $L$ is
large  $L\gg H^{-1}$, we have a large number of KK modes which contribute.
The effect of these fields on tensor fluctuations is shown to be
small. 
In our approach, the enhancement of scalar perturbation to
tensor perturbation is due to the large 
number of fields that contribute to the former. 

In this paper we first compute fluctuations assuming the background
is pure de Sitter, and later discuss the changes needed when Hubble 
is time dependent. Since the fluctuations originate from
massive fields, 
the spectrum is tiled towards the UV (spectral index $n_{s}>1$), 
if Hubble parameter were constant. However, the spectral
index is strongly 
dependent on the time-dependence of $H$. It
can be lowered
if Hubble decreases with time. We cannot know the dynamics of
Hubble unless we know the origin of vacuum energy during inflation. 
In this paper, we do not  make definitive statement, but we
mention the possibility that quantum fluctuations of these fields
(renormalized expectation value of energy momentum tensor) 
is the source of vacuum energy. 

Related work has been done by Nambu and Sasaki~\cite{Sasaki}. 
They computed correlation functions of energy-momentum tensor 
at the quadratic order in fluctuations, and related them to 
curvature perturbations. Their analysis is very similar to ours, 
but the setup and the interpretation are different. 
They consider a scalar field in an unstable potential $m^2<0$ 
(with a suitable regularization). Their goal is to rederive 
density fluctuations in slow-roll inflation from purely
quantum analysis without directly using the classical 
solution which rolls down the potential. On the other hand,
we are considering fields in the stable potential $m^2>0$,
and studying their vacuum fluctuations.

\subsection{Organization of this paper}

We will include descriptions of some known facts 
to make this paper self-contained
and to clarify our assumptions. 

In Section 2, we review quantization in de Sitter background. 
In Section 3, we study Einstein equations and express gravitational 
potential $\Phi$ in terms of matter fields. In Section 4, we obtain 
two-point functions of $\Phi$.
In Section 5, we find the CMB temperature fluctuations, and 
compare our formula with the
observed amplitude to find typical value of parameters
of fundamental theory.
In Section 6, we study the spectral index, and
discuss the effect of time-dependent Hubble constant. 
In Section 7, we consider non-Gaussianities. We compute 
three-point functions at the lowest order in the interaction, 
and we estimate the importance of interactions. 
In Section 8, we give a summary. 
In Appendix A, we perform the analysis of fluctuations
including an inflaton field as an effective model for time-dependent Hubble.

Part of the results of this paper has been reported in
our previous publication~\cite{letter}.

\section{Quantization in de Sitter space}

In this and the following two sections, we derive the formulas assuming
the background is pure de Sitter space. We will discuss
later what kind of changes are needed when Hubble is time
dependent. We start by reviewing the calculation of correlation functions
in de Sitter space, 
paying attention to the behavior in the small mass limit,
which will be important for later applications.

We will consider free fields, since we are mainly interested in
weakly coupled theories. The magnitude of temperature fluctuations 
described in this paper depends on the number of fields
which have masses smaller than the Hubble scale, but not on the 
details of the theory, so our conclusions will be valid even
in the presence of interactions.
We will discuss the effect of interaction in Sec. 7.

The metric of de Sitter space is 
\begin{align}
 ds^2 &=dt^2-a^2(t)d\vec{x}^2, \quad a(t)=H^{-1}e^{Ht},
\quad (-\infty\le t\le \infty)\\
&=a^2(\tau)\left(d\tau^2-d\vec{x}^2\right), \quad a(\tau)={1\over
 (-H\tau)}, \quad(-\infty\le \tau\le 0)
\label{deSitter}
\end{align}
where the conformal time $\tau$ is defined by 
$\tau=\int dt/a(t)=-e^{-Ht}$.

\subsection{Scalars}
Let us consider a free massive minimally-coupled scalar
field,
\begin{equation}
S={1\over 2}\int d^{4}x \sqrt{-g}\left\{\partial_{\mu}\phi\partial^{\mu}\phi
-m^2\phi^2 \right\}.
\end{equation}
It is convenient to define a rescaled field 
$\chi(\tau, \vec{x})=a(\tau)\phi(\tau, \vec{x})$ which
has the standard kinetic term.  
The equation of motion for the Fourier mode $\chi_{\vk}(\tau)$ 
where $\chi(\tau, \vec{x})
=\int {d^3k\over (2\pi)^3} \chi_{\vk}(\tau) e^{i\vec{k}\dot \vec{x}}$
is 
\begin{equation}
\chi_{\vec{k}}^{\prime \prime}(\tau)+\left\{|\vec{k}|^2
+\left(H^{-2}m^2-2\right)\frac{1}{\tau^2}\right\}
\chi_{\vec{k}}(\tau)=0.
\label{scalareom}
\end{equation}
Canonical quantization condition is
$[\chi(\tau,\vx), \chi'(\tau,\vx')]=i\delta^3(\vx-\vx')$.
We define the creation and annihilation operators $a^\dagger_\vk$, 
$a_\vk$ by 
\begin{equation}
\chi (\tau ,\vec{x})
=\int {d^{3}k\over (2\pi )^{3/2}}{1\over \sqrt{2|\vec{k}|}}
\left[u_{\vec{k}}(\tau )a_{\vec{k}}e^{i\vec{k}\cdot \vec{x}}
+u_{\vec{k}}^{\ast}(\tau )a^{\dagger}_{\vec{k}}e^{-i\vec{k}\cdot \vec{x}}\right]
\label{chimode}
\end{equation}
where $u_{\vec{k}}(\tau)$ is the solution of (\ref{scalareom}) 
which is normalized as $u_\vk\dot{u}_\vk^* -u_\vk^*\dot{u}_\vk
=2i|k|$. We take the solution which approaches
$u_{\vec{k}}(\tau)\to e^{-i|\vec{k}|\tau}$ at early time
$\tau\to-\infty$, so that the choice of the vacuum reduces to
the one for flat spacetime in the short-distance limit. 
We will take Bunch-Davies vacuum, which is annihilated by
$a_{k}$'s in (\ref{chimode}), throughout this paper. 
The explicit form of $u_\vk(\tau)$ is 
\begin{equation}
u_{\vec{k}}(\tau) =\sqrt{\pi\over 2}e^{i{\pi\over 2}(\nu+{1\over 2})}
\sqrt{-|\vec{k}|\tau} H_{\nu}^{(1)}(-|\vec{k}|\tau)
\end{equation}
with 
\begin{equation}
\nu=\sqrt{{9\over 4}-m^{2}H^{-2}}. 
\end{equation}
Asymptotic behavior at the late times (in the super-horizon 
$|\vec{k}|/a\ll H$ limit) 
is given by $u_{\vec{k}}\sim (-|\vec{k}|\tau)^{-\nu+{1\over 2}}$,
or in terms of the original field, $\phi\sim (-\tau)^{3/2-\nu}$,
as we can easily from the equation of motion (\ref{scalareom}):
The $|k|^2$ term drops out from the equation at late times, 
and the scaling 
w.r.t.\ time is independent of $|k|^2$;
de Sitter symmetry tells us that the spatial ($|k|$) dependence enters 
as a multiplicative factor with the same scaling dimension
as the one for $\tau$ 
(see e.g.~\cite{Strominger}).

Fields with small mass, 
\begin{equation}
mH^{-1}<{3\over 2},
\end{equation}
do not oscillate in time. The friction due to the cosmic 
expansion overdamps the oscillation due to energy of 
massive field. 

We are interested in correlation functions, which are the expectation
values taken with Bunch-Davies vacuum as in and out state.  
Two-point function at equal time is given by 
\begin{equation}
 \langle \phi(\tau, \vec{x} )\phi(\tau, \vec{x}' )\rangle
= {1\over a^{2}(\tau)}\int {d^{3}k\over (2\pi)^3}{1\over 2|k|}
|u_k(\tau)|^{2}e^{i\vec{k}(\vec{x}-\vec{x'})}.
\end{equation}
Substituting the late time expression for $u_{k}(\tau)$, we get
\begin{equation}
\langle \phi (\tau, x) \phi (\tau, x')\rangle
=H^{2}C(\gamma)(Ha|x-x'|)^{-\gamma}
\end{equation}
where 
\begin{equation}
 \gamma=3-2\nu,
\label{gamma}
\end{equation}
and
\begin{equation}
 C(\gamma)={\sin ({\pi\over 2}(\gamma-1))\Gamma({\gamma\over 2}) 
\over 4 \pi^{3/2} \{\sin({\pi\over 2}(3-\gamma))\}^{2}
\Gamma({\gamma-1\over 2})}.
\end{equation}
In the limit of small mass $mH^{-1}\ll 1$, we have 
$\gamma\sim {2\over 3}m^{2}H^{-2}$, and
\begin{equation}
 C(\gamma)\sim {1\over 4\pi^{2}\gamma}.
\end{equation}
The coefficient $C(\gamma)$
diverges in the $mH^{-1}\to 0$ limit, but
physical quantities such as the gravitational potential $\Phi$
stays finite in this limit as we will see below. 
In the massless limit, 
the exponent $\gamma$ approaches zero, and the decay
is slowest. We will see that fields with small mass
(more precisely $mH^{-1}\lesssim 10^{-1}$) mostly 
contribute to $\Phi$. 

The energy-momentum tensor for a minimally coupled scalar is
given by
\begin{equation}
 \delta T_{\mu\nu}=\left\{ \partial_{\mu}\phi\partial_{\nu}\phi
-{1\over 2}g_{\mu\nu}(\partial^{\rho}\phi \partial_{\rho}\phi
-m^{2}\phi^2)\right\}.
\end{equation}
This serves as the source for the gravitational fields.
Let us look at the time-dependence of the $\delta T_{00}$ 
component. Since $\phi\sim (-\tau)^{\gamma/2}$, the leading
term of $\delta T_{00}$ scales as 
\begin{equation}
\delta T_{00}\sim (-\tau)^{\gamma-2}.
\label{gamma2} 
\end{equation}
We will see in the next section that this produces the 
gravitational potential $\Phi\sim (-\tau)^{\gamma}$, which 
decays slowly when $\gamma\sim {2\over 3}m^{2}H^{-2}\ll 1$.
The fields that give important contributions are those which
give $\delta T_{00}\sim (-\tau)^{-2+O(m^2H^{-2})}$ at late
times. We can safely neglect the fields for which 
$\delta T_{00}$ decay faster than this, when we compute $\Phi$
in the late time limit. 

So far we have considered minimally-coupled scalar. 
If there is coupling to the curvature, the action becomes
\begin{equation}
 S={1\over 2}\int d^{4}x \sqrt{-g}\left\{\partial_{\mu}\phi\partial^{\mu}\phi
-(m^2+\xi R)\phi^2 \right\}.
\end{equation}
where $R$ is the scalar curvature of the background; $\xi(\ge 0)$ is
a constant, which takes the value $\xi=1/6$ for the conformally invariant
coupling. For de Sitter space, we have $R=12H^2$. The curvature 
coupling effectively increases the mass by
\begin{equation}
 m^{2}\to m^{2}+12H^2 \xi,
\end{equation}
and changes $\nu$ to
\begin{equation}
\nu=\sqrt{{9\over 4}-12\xi -m^{2}H^{-2}}.
\end{equation}

The late time behavior of such a field is
\begin{equation}
 \phi\sim (-\tau)^{{3\over 2}-\sqrt{{9\over 4}-12\xi -m^{2}H^{-2}}}.
\end{equation}
For the conformal scalar ($\xi=1/6$,  $m=0$), 
$\phi$ decays as $\phi\sim (-\tau)^{1}$. 
To have the exponent
close to zero so that the field contributes to $\Phi$, 
we need $\xi=0$ (minimal coupling) or close to zero,
and $mH^{-1}\ll 1$.

\subsection{Vectors}

Massive vector field (Proca field) is described by the
action
\begin{equation}
 S=\int d^{4}x\sqrt{-g}\left({1\over 4}g^{\mu\mu'}g^{\nu\nu'}
F_{\mu\nu}F_{\mu'\nu'}-{m^{2}\over 2}g^{\mu\mu'}
A_{\mu}A_{\mu'}\right).
\end{equation}
Vector field arising from the KK reduction of a gauge
field in higher dimension is such an example.

The equation of motion in de Sitter space (in the conformal coordinates)
is
\begin{equation}
 \eta^{\mu\mu'}\partial_{\mu}(\partial_{\mu'}A_{\nu}
-\partial_{\nu}A_{\mu'})+m^{2}a^{2}A_{\nu}=0.
\label{Procaeom}
\end{equation}
We act
$\partial_{\nu}$ on this equation, and find a constraint,
\begin{equation}
 \partial_{0}A_{0}+2{\cal H}A_{0}=\partial_{i}A_{i},
\end{equation}
where 
\begin{equation}
 \CH={a'\over a}=-{1\over \tau}.
\end{equation}

To solve the equation of motion, we decompose $A_{i}$
into the transverse and the longitudinal part,
\begin{equation}
 A_{i}=A_{i}^{(T)}+\partial_{i}\alpha
\end{equation}
where $\partial_{i}A^{(T)}_{i}=0$. The transverse part
satisfies
\begin{equation}
(\partial_{0}^{2}-\Lap +m^{2}a^{2})A_{i}^{(T)}=0.
\label{eomAi}
\end{equation}
The component $A_{0}$ satisfies the same equation after
a rescaling by the scale factor,
\begin{equation}
 (\partial_{0}^{2}-\Lap +m^{2}a^{2})(aA_{0})=0.
\label{eomA0}
\end{equation}
The scalar function $\alpha$ is determined by
\begin{equation}
 \Lap \alpha =\partial_{0}A_{0}-2{A_{0}\over \tau}.
\label{Lapalpha}
\end{equation}

These equations (\ref{eomAi}), (\ref{eomA0})
are equivalent to the equation of motion satisfied by
$a\varphi_{\rm conf}$ where $\varphi_{\rm conf}$ is a 
scalar with conformal coupling,
$m^{2}H^{2}\to m^{2}H^{2}+2$. At late time and in the 
limit of small mass, the fields scale as
\begin{align}
A_{0}&\sim \varphi_{\rm conf}\sim (-\tau)^{1+O(m^{2}H^{-2})}\\
A_{i}^{(T)}&\sim a \varphi_{\rm conf} 
\sim (-\tau)^{O(m^{2}H^{-2})}
\end{align} 
Also, from (\ref{Lapalpha}), we find $\alpha\sim (-\tau)^{O(m^{2}H^{-2})}$.

The energy-momentum tensor for massive vector field is
\begin{equation}
\delta T_{\mu\nu}=F_{\mu\rho}F_{\nu}{}^{\rho}-m^{2}A_{\mu}A_{\nu}
-{1\over 2}g_{\mu\nu}\left({1\over 2}F_{\rho\sigma}F^{\rho\sigma}
-m^{2}A_{\rho}A^{\rho}\right).
\end{equation}
To see the scaling of $\delta T_{00}$ at late times, 
let us look at its mass-dependent part,
\begin{equation}
\delta T_{00}^{(m^2)}=-{m^2\over 2}(A_{0}A_{0}+A_{i}A_{i}).
\end{equation}
The leading time-dependence is given by
\begin{equation}
 \delta T_{00}^{(m^2)}\sim -{m^2\over 2}a^2 \varphi_{\rm conf}^2
\sim (-\tau)^{2+O(m^{2}H^{-2})}. 
\end{equation}
The energy momentum tensor scales in the way as if we had
the scalar field $\varphi_{\rm conf}$. Whether the
field contribute to $\Phi$ or not depends on the effective
mass in the equation of motion. Vector 
fields in (3+1) dimensions have conformal coupling 
(due to the conformal invariance 
in the massless limit), and decay faster than the 
minimally coupled scalars at late times, so they do not 
contribute to $\Phi$.

\subsection{Spinors}

The action of Dirac spinor is 
\begin{equation}
 S=\int d^{4}x \sqrt{-g}\left\{\bar{\Psi}
\left(i e^{\mu}_{\hat{\mu}}\gamma^{\hat{\mu}}D_{\mu}-m\right)\Psi
\right\}.
\end{equation}
In the background (\ref{deSitter}), this is written as
\begin{equation}
 S=\int d^{4}x\left\{(a^{3/2} \bar{\Psi})
\left(i \gamma^{\hat{\mu}}\partial_{\hat{\mu}}-m a\right)(a^{3/2}\Psi)\right\},
\end{equation}
reflecting the fact that spinors are conformally invariant
in the massless case. 

Dirac equation is 
\begin{equation}
 (i\gamma^{\hat{\mu}}\partial_{\hat{\mu}}-ma)(a^{3/2}\Psi)=0, 
\end{equation}
and the independent components ($a^{3/2}\Psi=(\psi_{+}, \psi_{-})$
in certain representation of gamma matrices) satisfy
\begin{equation}
\left(\partial_{0}^{2}-\Lap +{m^{2}H^{-2}\pm imH^{-1}\over \tau^{2}}
\right)\psi_{\pm}=0.
\label{spinorpm}
\end{equation}
Note that in the massless limit, (\ref{spinorpm}) is the equation 
satisfied by the conformal scalar $a\varphi_{\rm conf}$. Thus,
the original field scales as $\Psi\sim a^{-1/2}\varphi_{\rm conf}\sim
(-\tau)^{3/2+O(m^2H^{-2})}$ at late times. 

Energy momentum tensor for spinors is (see e.g. \cite{BirrellDavies})
\begin{equation}
 \delta T_{\mu\nu}={i\over 2}\left\{\bar{\Psi}\gamma_{(\mu}D_{\nu)}\Psi
-(D_{(\mu}\bar{\Psi})\gamma_{\nu)}\Psi \right\}.
\end{equation}
The $\delta T_{00}$ component scales as 
\begin{equation}
 \delta T_{00}\sim \bar{\Psi}\gamma_{0}D_{0}\Psi
\sim (-\tau)^{1+O(m^2H^{-2})},
\end{equation}
since $\gamma_{0}=e_{\hat{\mu}0}\gamma^{\hat{\mu}}$ has one factor of $a\sim (-\tau)^{-1}$,
and $\partial_{0}$ decreases the power of $\tau$ by 1. 
This $\delta T_{00}$ is smaller than that for the
massless minimally coupled scalar, so spinors do not
contribute to $\Phi$ at late times. 

\subsection{The fields that are important at late times}

We have seen that minimally coupled scalar with mass
$mH^{-1}\ll 1$ decays most slowly, $\phi\sim(-\tau)^{O(m^2H^{-2})}$,
in the late time limit. Coupling to the curvature effectively
increases the mass, and fields such as conformal scalars
decay faster. 
The fields whose independent components scale in the
same way as minimally coupled scalar can contribute to 
$\Phi$ in the late time limit.

We can have small mass for the KK modes when extra dimensions
are large enough $L\gg H^{-1}$. Let us list possible origins of 
the fields which have minimal coupling.
\begin{itemize}
\item Massless minimally coupled scalars in
higher dimensions.
\item The scalar fields from the 
KK reduction of gauge fields with indices
along the internal directions: As long as the
size of the extra dimension is stabilized independently
of the scale factor for the 4D spacetime, these field do
not have coupling to the curvature. Higher dimensional
graviton with indices in the internal directions
is also such an example.
\item Massive tensors (in 4D) from the KK reduction of 
higher dimensional gravitons: In the massless 
limit, the transverse mode satisfy the equation 
of motion equivalent to massless minimally coupled
scalar (see (\ref{tensoreq}) below), thus contributes
at late times.    
\end{itemize}
Whether the first type of fields exist or not may
depend on the theory, but the second and the third
(one-form gauge fields and gravitons) 
will exist in fundamental theories in general. 
In the following, we will not ask how many of
these fields exist. We will ignore order 1 factor
coming from this multiplicity, since this is much
smaller than the huge multiplicity of the
KK modes for each field. In the explicit analysis, 
we will take minimally coupled scalar fields. 
Other fields can be studied in the similar
manner by considering the independent components 
which satisfy scalar-type equations of motion,
as long as we are considering vacuum fluctuations
of these fields.

\section{Einstein equations}
We now study Einstein equations. Einstein equations are
constraint equations which allow us to write the gauge
invariant metric fluctuations (such as gravitational 
potentials $\Phi$ and $\Psi$) in terms of matter fields. 
The metric fluctuations are decomposed into scalar, vector,
and tensor modes, each of which can be studied separately. Tensor
mode is the part which is transverse-traceless in 
the spatial directions, vector modes are those which are
divergenceless, and scalar modes are those which can
be written as derivatives of scalar functions. We 
follow the notation of \cite{Mukhanov}.

\subsection{Scalar fluctuations}
The scalar part of the (0,0), (0, $i$), ($i$, $j$) components of 
Einstein equations are given, respectively, by
\begin{eqnarray}
&&\Lap\Psi-3\CH (\Psi'+\CH\Phi)=4\pi G\delta T_{00},
\label{E00}\\
&&(\Psi'+\CH\Phi),{}_{i} =4\pi G\delta T^{\rm (S)}_{0i},
\label{E0i}\\
&&\left[\Psi''+\CH (2\Psi+\Phi)'+(2\CH'+\CH^2)\Phi
+{\Lap\over 2}(\Phi-\Psi)\right]\delta_{ij}
-{1\over 2}(\Phi-\Psi),{}_{ij}=4\pi G\delta T^{\rm (S)}_{ij}.
\nonumber\\
\label{Eij}
\end{eqnarray}
We take the background spacetime to be pure de Sitter space
($a=-H^{-1}/\tau$, $\CH ={-1/\tau}$). 
The l.h.s.\ are the Einstein tensor expanded to the 1st order
in metric fluctuations. $\Phi$ and $\Psi$ are the two gauge invariant
variables constructed from the scalar components. 
In the longitudinal gauge, they are given by 
\begin{equation}
 ds^{2}_{\rm l. g.}=a^2\left\{(1+2\Phi)d\tau^2-(1-2\Psi)\delta_{ij}
dx^{i}dx^{j}\right\}.
\end{equation}

On the r.h.s., we take the energy momentum tensor which is
quadratic in the matter fields. We consider 
minimally-coupled scalars here. 
We assume there are many free scalar fields.
The energy momentum tensor is a sum over their contributions,
\begin{equation}
 \delta T_{\mu\nu}=\sum\left\{ \partial_{\mu}\phi\partial_{\nu}\phi
-{1\over 2}g_{\mu\nu}(\partial^{\rho}\phi \partial_{\rho}\phi
-m^{2}\phi^2)\right\}.
\end{equation}
For brevity, the label on the field is suppressed, and
the sum is understood to be over the species.
The fields $\phi$ have vanishing classical background,
and are gauge invariant.  
Each component of $\delta T_{\mu\nu}$ is given by 
\begin{eqnarray}
 \delta T_{00}&=&\sum
{1\over 2}\{\phi'{}^2+\partial_{i}\phi\partial_{i}\phi
+m^{2}a^{2}\phi^2\},\\
\delta  T_{0i}&=&\sum\left\{\phi'\partial_{i}\phi\right\},\\
\delta  T_{ij}&=&\sum\left\{\partial_{i}\phi \partial_{j}\phi
+{1\over 2}\delta_{ij}(\phi'{}^2-\partial_{i}\phi\partial_{i}\phi
-m^2 a^{2}\phi^2)\right\}. 
\end{eqnarray}
The superscript ${\rm (S)}$ in (\ref{E0i}), (\ref{Eij}) denotes 
the scalar part. 
Recall that $\delta T_{0i}$ and $\delta T_{ij}$ can be 
decomposed as 
\begin{eqnarray}
\delta T_{0i}&= &\partial_{i}\tilde{s} + u_{i},
\label{T0i}\\
\delta T_{ij}&=&\partial_{i}\partial_{j}s
-{1\over 3}\delta_{ij}\Lap s +\partial_{i}v_{j}+\partial_{j}v_{i}+t_{ij}
+ f \delta_{ij},\label{Tij}
\end{eqnarray}
where $u_{i}$  and $v_{i}$ are transverse vectors $\partial_{i}v_{i}
=\partial_{i}u_{i}=0$, and $t_{ij}$ is a transverse traceless (TT)
tensor, $\partial_{i}t_{ij}=t_{ii}=0$. 
By the scalar part, we mean the part involving $\tilde{s}$ in
(\ref{T0i}), and the part involving $s$ and $f$ in (\ref{Tij}). 

We can find $\tilde{s}$ and $s$ by taking divergence and 
applying inverse Laplacian, 
\begin{equation}
\tilde{s}={1\over \Lap}\partial_{k} \delta T_{0k}
=\sum{1\over \Lap}\partial_{k}\left(\phi'\partial_{k}\phi\right),
\end{equation}
\begin{equation}
s={3\over 2\Lap^2}\partial_{k}
\partial_{l}(\delta T_{kl}-{1\over 3}\delta_{kl}\delta T_{mm})
=\sum {3\over 2\bigtriangleup^2}\partial_{i}\partial_{j}
(\partial_{i}\phi\partial_{j}\phi-{\delta_{ij}\over 3}
\partial_{k}\phi\partial_{k}\phi).
\label{TSij}
\end{equation}

Using Einstein equations (\ref{E00})-(\ref{Eij}), we can solve for
$\Phi$ and $\Psi$ in terms of $\phi$. 
First, from the traceless part of (\ref{Eij}),
we find
\begin{equation}
 \Phi-\Psi=-8\pi G s.
\label{PhiPsi}
\end{equation}
Using this in (\ref{E0i}), 
\begin{equation}
 \Phi'+\CH \Phi=  8\pi G \left\{
-s' +{1\over 2\Lap}\partial_{i}
(\phi' \partial_{i}\phi)\right\}.
\label{E0i2}
\end{equation}
The last term is the part that we would get if we had $\Phi=\Psi$. 

Let us solve (\ref{E0i2}) by substituting the late time asymptotics
of $\phi$ on the r.h.s. This is a valid procedure, since we are 
interested in the correlation
functions of $\Phi$ in the late time limit, and $\Phi$ only appears
as external lines. Special care is needed
if the leading term (which has the lowest scaling dimension) is 
degenerate with another term, which can happen at certain values
of the parameter; we will comment on this point when necessary.

In the late time limit, the time- and space- dependence of 
the field $\phi$ factorizes,
\begin{equation}
 \phi(\tau, x)=(-\tau)^{\gamma\over 2}\hat{\phi}(x),
\end{equation}
where $\gamma$ is defined in (\ref{gamma}).
Time-dependence of $\Phi$ is found from the time-dependence of the
r.h.s.\ of (\ref{E0i2}),
\begin{equation}
\tau ({1\over \tau}\Phi)' \sim (-\tau)^{\gamma-1}\quad
 \Rightarrow \quad \Phi\sim (-\tau)^{\gamma}.
\end{equation} 
Thus, $\Phi$ at late times can be written as
\begin{equation}
 \Phi(\tau, x)=(-\tau)^{\gamma}\hat{\Phi}(x).
\end{equation}
Time derivative is given by   $\Phi'={\gamma\over \tau} \Phi$,
$\Psi'={\gamma\over \tau} \Psi$, $\phi'={\gamma\over 2\tau} \phi$.
From (\ref{E0i2}), we get
\begin{equation}
\Phi=4\pi G {\gamma\over \gamma-1} \left\{-{3\over \Lap^2}
\partial_{i}\partial_{j}(\partial_{i}\phi 
\partial_{j}\phi)
+{1\over \Lap}\partial_{i}\phi\partial_{i}\phi
+{1\over 4}\phi^2\right\}, 
\label{solPhi}
\end{equation}
where we have used $\partial_{i}(\phi\partial_{i}\phi)
=\Lap\phi^2/2$ to rewrite the last term. 
This solution is consistent with all the other components of
Einstein equations.

The expression (\ref{solPhi}) diverges at $\gamma=1$. At this
special value, (\ref{E0i2}) cannot be solved with the naive 
ansatz $\Phi\sim (-\tau)^{\gamma}$, since the l.h.s.\  vanishes.
In this case, we can solve the equation by setting 
$\Phi\sim (-\tau)\log(-\tau)$.

Also note that there is always a freedom of adding a term which 
has time-dependence  $\sim (-\tau)^{1}$ to the solution 
of (\ref{solPhi}), but we can eliminate this piece by
requiring that the solution does not blows up 
in the early time limit.

\subsection{Vector fluctuations}
The vector modes are the following part of the metric fluctuations,
\begin{equation}
 ds^{2}=a^{2}\left[ d\tau^2 +2S_{i}dx^{i}d\tau
-(\delta_{ij}-F_{i,j}-F_{j,i})dx^{i}dx^{j}\right],
\end{equation}
where 
$\partial_{i}S_{i}=\partial_{i}F_{i}=0$, and
$,i$ denotes derivative w.r.t.\ $x^i$.
There is a gauge invariant combination,
\begin{equation}
 V_{i}=S_{i}-F'_{i}.
\end{equation}
The vector part of the (0, $i$) and ($i$, $j$) components of the 
Einstein equations are 
\begin{align}
&\Lap V_{i}=16\pi G\delta T^{(V)}_{0i},
\label{Vi}\\
&(V_{i,j}+V_{j,i})'+2\CH (V_{i,j}+V_{j,i})
=16\pi G\delta T_{ij}^{(V)},
\label{Vij}
\end{align}
where the superscript $(V)$ denotes the vector part,
which are the part of $\delta T_{0i}$ and $\delta T_{ij}$
which involves  $u_{i}$ and $v_{i}$, as defined 
in (\ref{T0i}) and (\ref{Tij}). 
$u_{i}$ is given by subtracting the scalar part from
$\delta T_{0i}$,  
\begin{equation}
\delta T_{0i}^{(V)}= u_{i}=\phi' \partial_{i}\phi -{1\over \Lap}\partial_{i}
\partial_{k}(\phi'\partial_{k}\phi).
\label{ui}
\end{equation}

The leading term of $u_{i}$ at late times is smaller than
it naively looks. Recall that $\phi\sim (-\tau)^{\gamma/2}
\hat{\phi}(1+O(\tau^2))$. We can see that the order 
$(-\tau)^{\gamma-1}$ term of the r.h.s.\ of (\ref{ui}) 
vanishes by using $\phi'={\gamma\over 2\tau}\phi$ and
$\phi\partial_{i} \phi=\partial_{i}(\phi^{2})/2$. Thus
the leading term of the r.h.s.\ of (\ref{Vi}) scales as 
$(-\tau)^{\gamma+1}$,
which implies $V_{i}\sim (-\tau)^{\gamma+1}$.
This behavior is consistent with the equation (\ref{Vij}).

Since $V_{i}$ decays at least as $(-\tau)^{1}$, we conclude
that the vector perturbation produced by the matter fields
$\phi$ is negligible at late times.

\subsection{Tensor fluctuations}
The transverse-traceless (TT) tensor fluctuation $h_{ij}$ 
($\nabla^{i}h_{ij}$=$h^{i}_{i}=0$) is defined by
\begin{equation}
 ds^{2}=a^{2}\left[d\tau^2-(\delta_{ij}-h_{ij})dx^{i}dx^{j}\right].
\end{equation} 
It is sourced by the
TT part of energy-momentum tensor, 
\begin{equation}
 h''_{ij}+2\CH h'_{ij}-\Lap h_{ij}=8\pi G \delta T^{(T)}_{ij}.  
\label{tensoreq}
\end{equation}

The general solution to this equation is given by
the solution $h^{(0)}_{ij}$ for the homogeneous equation
on top of a particular solution $h^{(1)}_{ij}$
which depends on $\delta T^{(T)}_{ij}$.

The homogeneous equation is equivalent 
to massless scalar equation of motion. Its solution $h^{(0)}_{ij}$
is the usual gravitational wave, which scales 
logarithmically in space and time. This has the scale invariant 
spectrum with the amplitude $H/m_{pl}$. 

The time-dependence
of $h^{(1)}_{ij}$ is determined by (\ref{tensoreq}) to be  
$h^{(1)}_{ij} \sim (-\tau)^{\gamma+2}$,
since $\delta T^{(T)}_{ij}\sim (-\tau)^{\gamma}$.
Care is needed when $\gamma=1$. In this case, $(-\tau)^{\gamma+2}
=(-\tau)^{3}$ is degenerate with the (decaying) solution of the
homogeneous equation, and (\ref{tensoreq}) cannot be solved
with this ansatz. In this case we have to take $h^{(1)}_{ij}\sim
(-\tau)^{3}\log(-\tau)$. 
In any case, $h^{(1)}_{ij}$ decays at late times, and the
effect of $\delta T^{(T)}_{ij}$ for the tensor fluctuations
is negligible at late times.

\section{Correlation functions}

Having expressed $\Phi$ in terms of $\phi$, it is straightforward
to compute correlation functions of $\Phi$.
Let us compute the two-point function.

We decompose $\Phi$ in (\ref{solPhi}) into two pieces,
\begin{eqnarray}
 \Phi&=&\Phi_0+\Phi_1\\
\Phi_{0}&=&-\pi G{\gamma\over 1-\gamma}\phi^2
\label{Phi0}\\
\Phi_{1}&=&4\pi G{\gamma\over 1-\gamma}\left({3\over \Lap^{2}}
\partial_{i}\partial_{j}(\partial_{i}\phi\partial_{j}\phi)
-{1\over \Lap}(\partial_{i}\phi\partial_{i}\phi)\right),
\label{Phi1}
\end{eqnarray}
where $\Phi_0$ is the part which we would get when $\Phi=\Psi$,
and $\Phi_1$ is the part which depends on $s$ defined in (\ref{TSij}).

The $\langle \Phi_{0}\Phi_{0}\rangle$ correlator is
just a product of two propagators, 
\begin{eqnarray}
 \langle \Phi_{0}(\tau, x)\Phi_{0}(\tau, x')\rangle
&=&2(\pi G)^{2}\sum \left({\gamma\over 1-\gamma}\right)^2
\langle \phi(\tau, x)\phi(\tau, x')\rangle^{2}\nonumber\\
&=&2(\pi GH^{2})^{2}\sum \left({\gamma\over 1-\gamma}\right)^2
C^{2}(\gamma)(-\tau)^{2\gamma}|x-x'|^{-2\gamma},
\label{Phi0Phi0}
\end{eqnarray}
where as in the last section, the sum is taken over all
species of $\phi$. When the mass of the fields are small
($\gamma\ll 1$), we get
\begin{equation}
  \langle \Phi_{0}(\tau, x)\Phi_{0}(\tau, x')\rangle
\sim {1\over 8\pi^{2}}(GH^{2})^{2}\sum
(-\tau)^{2\gamma}|x-x'|^{-2\gamma}.
\label{Phi0Phi0a}
\end{equation}
The contribution from the fields with $\gamma\ll 1$ is 
finite, since the two factors of $1/\gamma$ from the propagator
are canceled by the two factors of $\gamma$ from (\ref{Phi0}).  

The other parts of the correlator can be computed by using
the formulas such as
($\partial'_{i}={\partial\over \partial x'_{i}}$)
\begin{eqnarray}
 \partial_{i}\partial'_{j}{1\over |x-x'|^{\lambda}}
&=&\lambda\left\{{\delta_{ij}\over |x-x'|^{\lambda+2}}
-(\lambda +2){(x_{i}-x'_{i})(x_{j}-x'_{j})\over
|x-x'|^{\lambda+4}}\right\}\\
\Lap{1\over |x-x'|^{\lambda}}&=&{\lambda(\lambda-1)\over
|x-x'|^{\lambda+2}},
\end{eqnarray}
which are valid up to possible contact terms. The cross term 
$\langle \Phi_{1}\Phi_{0}\rangle$ is 
\begin{eqnarray}
\langle \Phi_{1}(\tau, x)\Phi_{0}(\tau, x')\rangle
&=&-8(\pi GH^{2})^{2}\sum \left({\gamma\over 1-\gamma}\right)^2
\Big[{3\over \Lap^{2}}\partial_{i}\partial_{j}
\langle \partial_{i}\phi(\tau, x)\phi(\tau, x')\rangle
\langle \partial_{j}\phi(\tau, x)\phi(\tau, x')\rangle\nonumber\\
&&\qquad -{1\over \Lap}\langle \partial_{i}\phi(\tau, x)\phi(\tau, x')\rangle
\langle \partial_{i}\phi(\tau, x)\phi(\tau, x')\rangle
\Big]\nonumber\\
&=&-4(\pi GH^{2})^{2}\sum \left({\gamma\over 1-\gamma}\right)^{2}
\left({\gamma\over 1+\gamma}\right)
C^{2}(\gamma)(-\tau)^{2\gamma}|x-x'|^{-2\gamma}.
\label{Phi0Phi1}
\end{eqnarray}

In the following, we will find that the fields that mostly
contribute to the CMB temperature fluctuations are the
ones with $mH^{-1}\lesssim 10^{-1}$, so let us study the
$\gamma\ll 1$ behavior here. In this limit, 
(\ref{Phi0Phi1}) is smaller than (\ref{Phi0Phi0a})
by a factor of $\gamma$. 

The part $\langle \Phi_{1}\Phi_{1}\rangle$,
\begin{eqnarray}
\langle \Phi_{1}(\tau, x)\Phi_{1}(\tau, x')\rangle
&=&16(\pi GH^{2})^{2}\sum \left({\gamma\over 1-\gamma}\right)^{2}
{\gamma^{2} (2\gamma^2+4\gamma-3)\over 
(\gamma+1)(\gamma+3)(2\gamma-1)(2\gamma+1)}\nonumber\\
&&\times C^{2}(\gamma)(-\tau)^{2\gamma}|x-x'|^{-2\gamma}.
\end{eqnarray}
is smaller than (\ref{Phi0Phi0a}) by a factor of $\gamma^2$.
Thus, we can consider only the $\langle \Phi_{0}\Phi_{0}\rangle$
part when $\gamma\ll 1$.

\subsection{Summing up KK modes}

Let us perform the summation over the massive fields $\phi$,
assuming they are Kaluza-Klein modes from the
compactification of extra dimensions. For definiteness,
let us assume there are $D$ dimensions which are
compactified on a torus $T^D$ with the periodicity $L$
(assumed to be the same for all directions for simplicity)
being large compared to the inverse Hubble of inflation,  
$L\gg H^{-1}$. We assume the internal directions other
than these $D$ dimensions are compactified on a space with 
the string scale size. 

The mass of a KK mode with $\{n_a\}$ units of momentum 
on $T^{D}$ is
\begin{equation}
m^2=\sum_{a=1}^{D} {(2\pi n_{a})^2\over L^2}.
\end{equation}
When the level is sufficiently dense, 
the density of states in the mass interval $dm$ around $m$ is
given by 
\begin{equation}
 S_{D-1}|n|^{D-1}d|n|=S_{D-1}(L/2\pi)^{D} m^{D-1}dm,
\label{dndm}
\end{equation}
where $S_{D-1}=2\pi^{D/2}/\Gamma(D/2)$ is the volume of
the $D-1$ dimensional unit sphere. This relation states that
the number of states is proportional to the phase 
space volume (the volume element of the KK momentum space 
times the volume of the internal space). 

Using (\ref{dndm}), we convert the sum in the 
$\langle\Phi\Phi\rangle$ correlator to an integral,
\begin{equation}
 \langle \Phi\Phi\rangle
=c_{D}L^{D}
 \left({H\over m_{pl}}\right)^{4}
\int_{0}^{m_{c}}dm m^{D-1} 
(Ha|\vec{x}-\vec{x'}|)^{-2\gamma},
\label{integral}
 \end{equation}
where  $c_{D}=S_{D-1}/(4(2\pi)^{D+2})$. 
We have used the expression (\ref{Phi0Phi0a}) for the correlator
in the $\gamma\ll 1$ limit.
The upper limit $m_{c}$ of the integration should be 
$m_{c}\sim {3\over 2}H$ as long as we are working in 
Einstein gravity, so that the field $\phi$ are the
ones which do not oscillate. 

\begin{figure}[htb]
\begin{center}
\includegraphics[height=5cm]{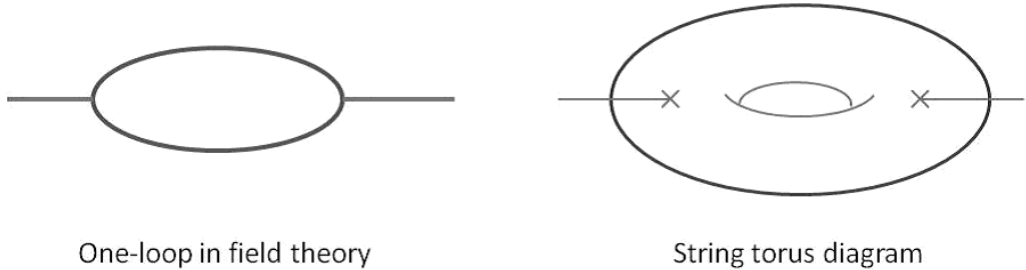}
\end{center}
\caption{One-loop diagram: One-loop diagram in field theory
corresponds to the torus diagram in string theory. 
String theory can be regarded as a field theory with an 
infinite number of fields, which are the Fourier modes on 
the world sheet spatial direction.}
\label{oneloop}
\end{figure}

\begin{figure}[htb]
\begin{center}
\rotatebox{-90}{
\includegraphics[height=10cm]{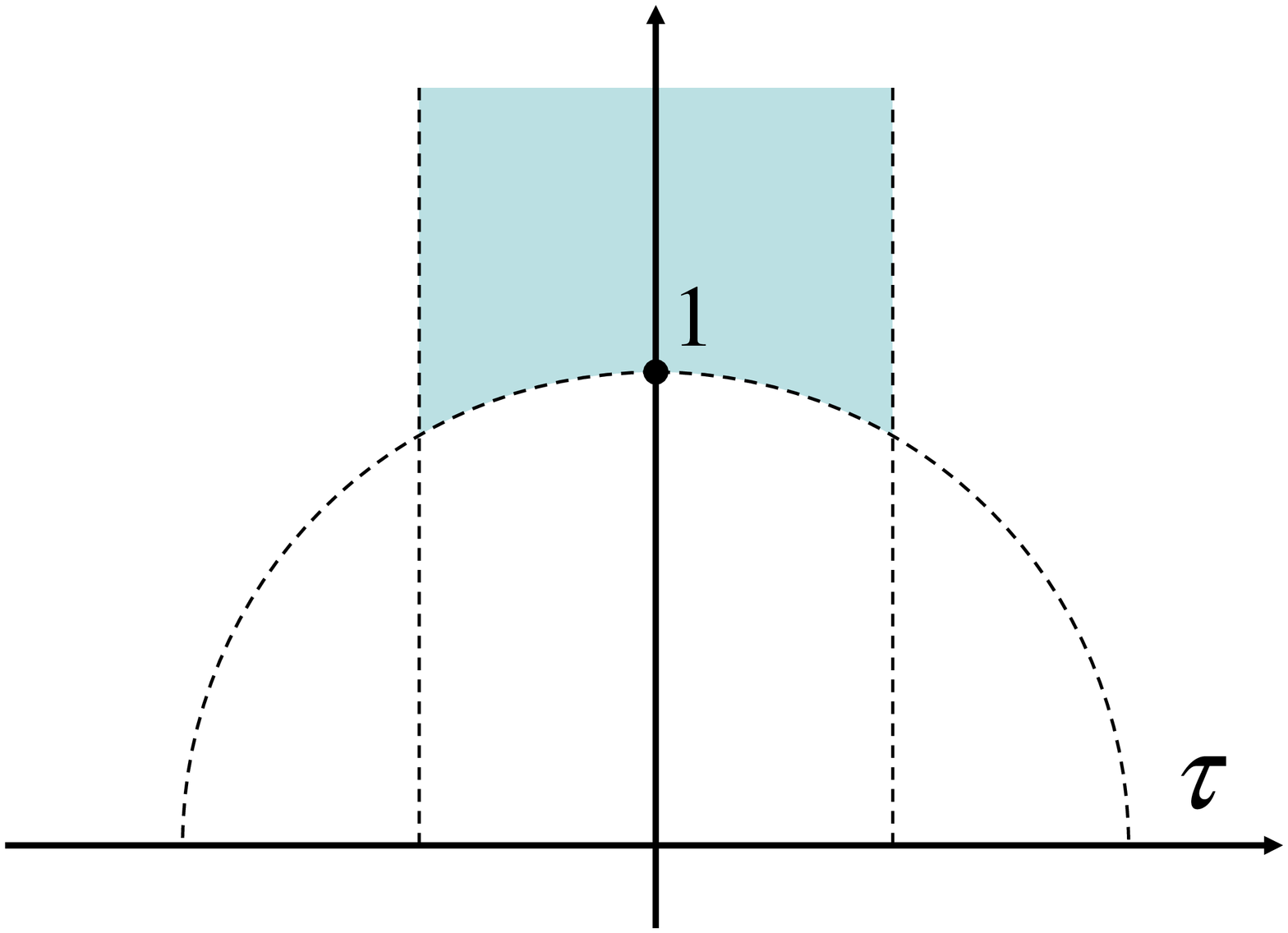}
}
\end{center}
\caption{Fundamental region: Integration over the moduli of the
torus should be restricted to the fundamental region, since
other regions are physically equivalent to this region due
to modular invariance. Im($\tau$), which corresponds to the
Schwinger parameter, is cut off at the distance of order string
scale.}
\label{fundamental}
\end{figure}

However, if string scale is less than Hubble scale, 
$m_{s}< H$, string states also have to be taken into account.
In this case, we expect that the sum over the mass 
is effectively cut off at $m_{c}\sim m_{s}$ for the following
reason. Let us assume the two-point function of $\Phi$ 
comes from the one-loop diagram in string theory 
(Fig~\ref{oneloop}).
String theory can be regarded as a field theory with infinitely 
many fields, except that one-loop amplitude effectively has 
UV cutoff due to modular invariance. The integral over the
moduli $\tau$ is restricted to the fundamental domain 
(Fig.~\ref{fundamental}), and the Schwinger proper 
time (Im($\tau$)) is cut off at string scale. There is no 
physical meaning
to time interval shorter than string scale, or oscillations much
higher than string scale. This means the internal states in the
loop which has mass much larger than string scale do not
have physical effect. This argument is based on the
perturbative string theory in flat spacetime, and it is not
clear whether it is valid in an arbitrarily curved background,
but we believe this is a reasonable estimate\footnote{In fact,
the precise value of the upper limit of integration is not
very important in the parameter region of interest. The conclusion 
that $m_sH^{-1}\sim 0.1$ is favored does not change even if we 
take the upper limit to be $(3/2)H$ instead of $m_{s}$.}.

\section{CMB temperature fluctuations}

\subsection{Generation of adiabatic fluctuation}

The formula (\ref{integral}), which was obtained with constant
$H$, is approximately valid during inflation. 
$H$ decreases towards the end of inflation. 
When $H$ becomes less than the mass of a field $\phi$, the
field will undergo classical oscillation, and decay into
radiations and stable particles. The energy density from
this process produces curvature perturbation. This is 
similar to what happens in 
the ``curvaton scenario''~\cite{curvaton},
in which the curvaton field (different from inflaton, 
and usually assumed to be one field), 
produces curvature perturbation. 
The situation that we are considering has 
similarities and differences with the situation 
usually considered in curvaton scenario. Even though it is 
helpful to have curvaton mechanism in mind, we emphasize 
the following specific assumptions that we make. 

We assume that the KK modes and string states decay 
sufficiently fast so that they do not interfere with
the standard big bang nucleosynthesis. The decay products
are assumed to be in local thermal equilibrium, 
and can be treated as a single fluid which dominates the
energy density of the universe. 
We do not assume low-energy (such as TeV scale) supersymmetry.
In such a case, it will be generically expected that the 
decay products interact among themselves, reaching thermal
equilibrium. This is the 
reheating mechanism in our model. The particles that
have conserved quantum numbers at present, such as
baryons, cold dark matter, etc.\ will be produced after
the above thermal equilibrium has been established.
In this case, fluctuations of the density of these species
are determined solely by the local temperature, and
obey the adiabaticity relation
(see e.g. discussion in \cite{Weinberg}). In the 
context of curvaton,
this corresponds to the case where matter (such as cold 
dark matter and baryons) are produced after curvaton 
decays, in which case there is no isocurvature 
fluctuations~\cite{curvaton2}. Even though the mass scale
of KK modes is lower than the Hubble of inflation, it will
be much higher than the scale of Standard Model of particle
physics. In such a case, we do not have reason to expect
that 
the fluctuations of particular species of KK modes to be
directly related to those of cold dark matter or baryons. 
Thus, we do not expect there to be  
significant isocurvature perturbation (which is strongly 
constrained by observation;
see e.g.\ Section 3.6 of \cite{WMAP5}) in our model.

The amplitude of the energy density fluctuations 
produced by the above process is determined by the quantum 
fluctuations of $\phi$'s during inflation. We assume the
transition from quantum fluctuations to classical oscillation 
and the decay of these fields occurs quickly and more or less 
simultaneously for all the fields.
(This is a simplification to make the argument simple; we
leave more general analysis to future work.) Under this
assumption, we identify $\Phi$ evaluated at the end of
inflation as the one sourced by the thermal fluid described
above.
Once the universe is in thermal equilibrium with a single
fluid dominating the density, 
the superhorizon mode of $\Phi$ basically remains 
constant (see e.g.\ \cite{Mukhanov}). It changes only by 
order 1 factor at the transition between matter and 
radiation domination, but ignoring this factor, $\Phi$ at 
the end of inflation is directly related to its value at the
recombination. 

One may worry that this mechanism produces anisotropy,
since there are fields with non-zero spin (such as KK modes of gravitons)
whose components separately undergo classical oscillation.
However, note that the inhomogeneities $\delta\rho/\rho$ that 
results from 
the fluctuations of many fields scales as $1/\sqrt{N}$ as the 
number of fields $N$ increases. This is because the possible classical 
density $\rho$ will be proportional to $N$, but the fluctuations 
$\delta\rho$ (or more precisely, the root of the square expectation value 
$\sqrt{\langle \delta\rho^2\rangle}$) will be of order $\sqrt{N}$ 
if the fields fluctuate independently. Therefore, total anisotropy
from the fluctuations of many fields scales as $1/\sqrt{N}$ and is 
kept small. This type of suppression of anisotropy 
due to a large number of fields appears also in the context
of ``vector inflation''~\cite{vector}.

\subsection{Amplitude of the CMB temperature fluctuations}

Having stated our assumptions which lead to the identification of
$\Phi$ at the end of inflation with $\Phi$ at recombination,
let us now study the CMB temperature fluctuations
$\delta T/T$. 

Temperature fluctuation of CMB is related to $\Phi$ at recombination
(at redshift $z\sim 1100$) by ${\delta T/ T} =-\Phi/3$. 
The angle $\theta$ on the sky corresponds
to the distance $d_{r}=2R_{r}\sin(\theta/2)$, where $R_r$ 
is the radius of the surface of last scattering. This is
of order the inverse of the present Hubble parameter
$R_{r}\sim H_{0}^{-1}$. 
The modes outside the horizon at recombination
correspond to the angle $3^{o}\le \theta$ (or angular
momentum $l\le 60$). 
These modes have been outside the horizon since the inflation, 
and $\Phi$ is frozen (remain constant at fixed comoving distance).
The distance $d_{r}$ corresponds to the distance 
\begin{equation}
 {(a_{e}/a_{r})}d_{r}=2 R\sin(\theta/2)=a_{e}|\vec{x}-\vec{x'}|
\end{equation} 
at the end of inflation. 
The radius $R=(a_{e}/a_{r})R_{r}$ will depend on the scale
of inflation.  We will take the standard estimate 
$RH\sim 10^{29}\sim e^{67}$ in the following, which amounts
to the assumption that the reheating temperature is not
much lower than the grand unification scale.

The angular power spectrum $C_{l}$ is defined by
\begin{equation}
 \langle {\delta T\over T}(\theta) {\delta T\over T}(0)\rangle
=\sum_{l=1}^{\infty} (2l+1) C_{l}P_{l}(\cos\theta).
\end{equation}
We will focus on the superhorizon modes. To find amplitude
of these modes, we expand the coordinates in 
the correlation function (\ref{integral}) (around $\theta=\pi$)
as 
\begin{equation}
(Ha|\vec{x}-\vec{x'}|)^{-2\gamma}
\sim (2RH)^{-2\gamma}(1-2\gamma\log(\sin(\theta/2)),
\label{expansion}
\end{equation}
and recall
that $-2\log(\sin(\theta/2))$ is $1/(l(l+1))$ in harmonic
space. From (\ref{integral}), we find the square 
amplitude $\delta^{2}_{T}\equiv l(l+1)C_{l}$, 
\begin{eqnarray}
\delta^{2}_{T}&=&{2\over 27}c_{D}{L^{D}\over H^{2}}
 \left({H\over m_{pl}}\right)^{4}
\int_{0}^{m_{s}}dm m^{D+1}(2RH)^{-{4\over 3}m^{2}H^{-2}}\nonumber\\
&=&{2\over 27}c_{D}\left({m_{s}\over H}\right)^{2}(Lm_{s})^{D} 
\left({H\over m_{pl}}\right)^{4}{\cal M}_{D}(\zeta_{0}).
\label{amplitude}
\end{eqnarray}
Note that the integrand is strongly suppressed at 
$m\gtrsim H/10$, due to the factor 
$(2RH)^{-{4\over 3}m^{2}H^{-2}}$. In (\ref{amplitude}), we have
taken the upper limit to be $m_{c}=m_{s}$. This is a good
approximation even when $m_{s}>H$, since the integral has 
little contribution from the region near the upper limit. 
We have defined
\begin{equation}
 {\cal M}_{D}(\zeta)=\int_{0}^{1}dt e^{-\zeta t^2}t^{D+1}, \quad
\zeta_{0}={4\over 3}{m_{s}^{2}\over H^{2}}\log(2RH).
\end{equation}

To see the qualitative behavior of $\delta^{2}_{T}$, it would be
helpful to note ${\cal M}_{D}(\zeta_{0})\sim \zeta_{0}^{-{D+2\over 2}}$
when $\zeta_{0}\gg 1$. In this limit, we have
$\delta^{2}_{T}\sim ({H\over m_{pl}})^{4}(LH)^{D}(\log(2RH))^{-{D+2\over 2}}$
up to constant factors. $\delta^{2}_{T}$ is enhanced when
extra dimensions are large, $(LH)^{D}\gg 1$,  
since many fields contribute to it. $\delta^{2}_{T}$ becomes small
if $\log (2RH)$ were larger due to the decrease of massive wave
function at large separation.

\subsection{Comparison with the data}

We will now use observational data~\cite{WMAP},
\begin{equation}
 \delta_{\rm T}\sim 2.6\times 10^{-5}, \quad 
r_{t/s}\lesssim 0.22,
\label{WMAP}
\end{equation}
to constrain the parameters in our model. This implies
${H\over  m_{pl}} =\sqrt{\frac{9\pi}{2}
\delta_{\text{T}}^2r_{\text{t/s}}}
\lesssim 0.81\times 10^{-4}$.
Let us first assume this inequality is saturated.
Then the amplitude (\ref{amplitude}) provides the relation between
the two parameters $m_s$ and $L$, or equivalently,  
between $m_s$ and the string coupling $g_s$, since $L$ is written as 
$(Lm_s)^D=8\pi^6g_s^2(m_{pl}^2/m_s^2)$. 

\begin{figure}[htb]
\begin{center}
\rotatebox{-90}{
\includegraphics[height=10cm]{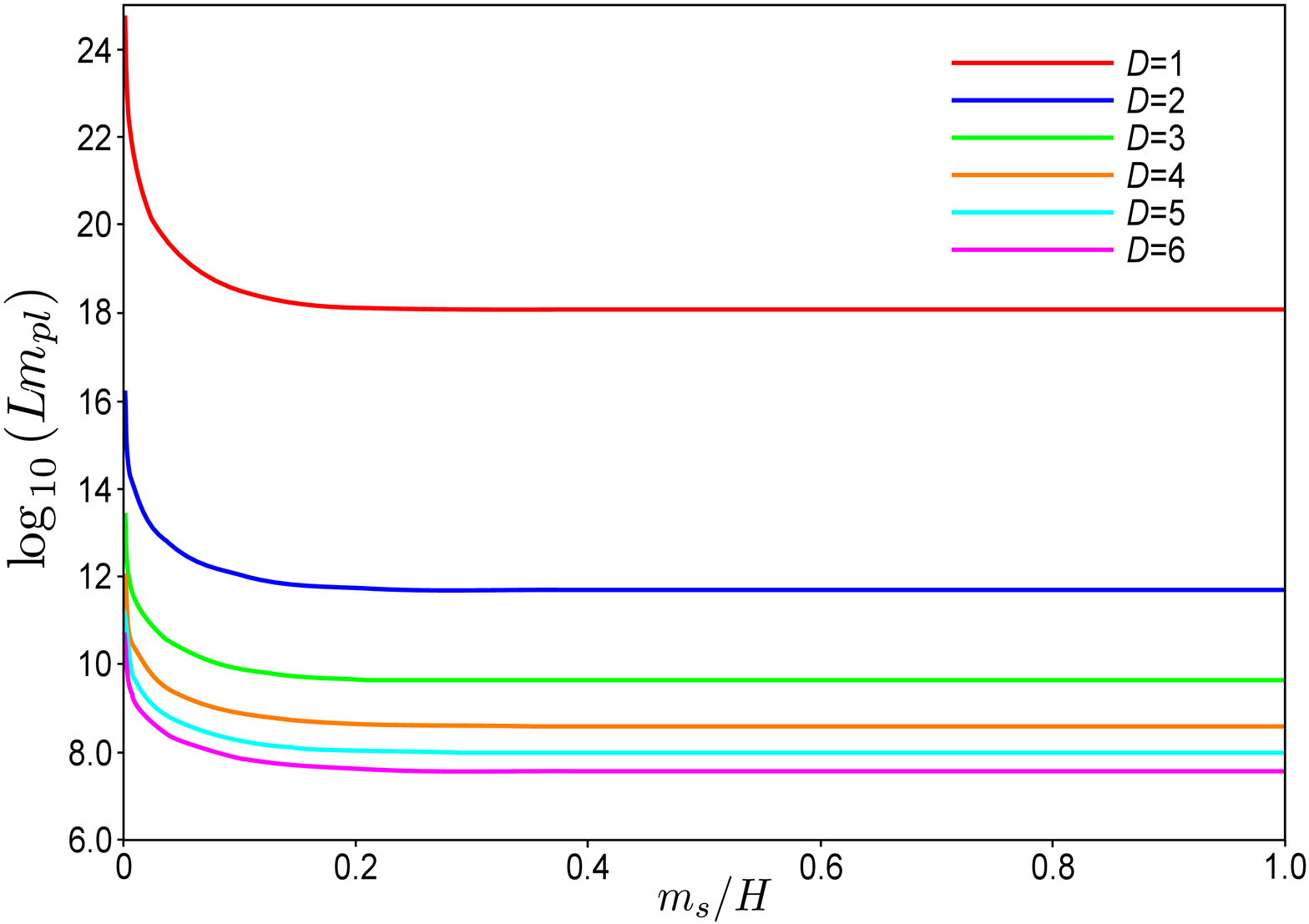}
}
\end{center}
\caption{$\log_{10}(Lm_{pl})$ as a function of 
$m_{s}/H$, with $\delta_{T}=2.6\times 10^{-5}$, 
$r_{\rm t/s}=0.22$, $RH\sim 10^{67}$. }
\label{Lms}
\end{figure}

\begin{figure}[htb]
\begin{center}
\rotatebox{-90}{
\includegraphics[height=10cm]{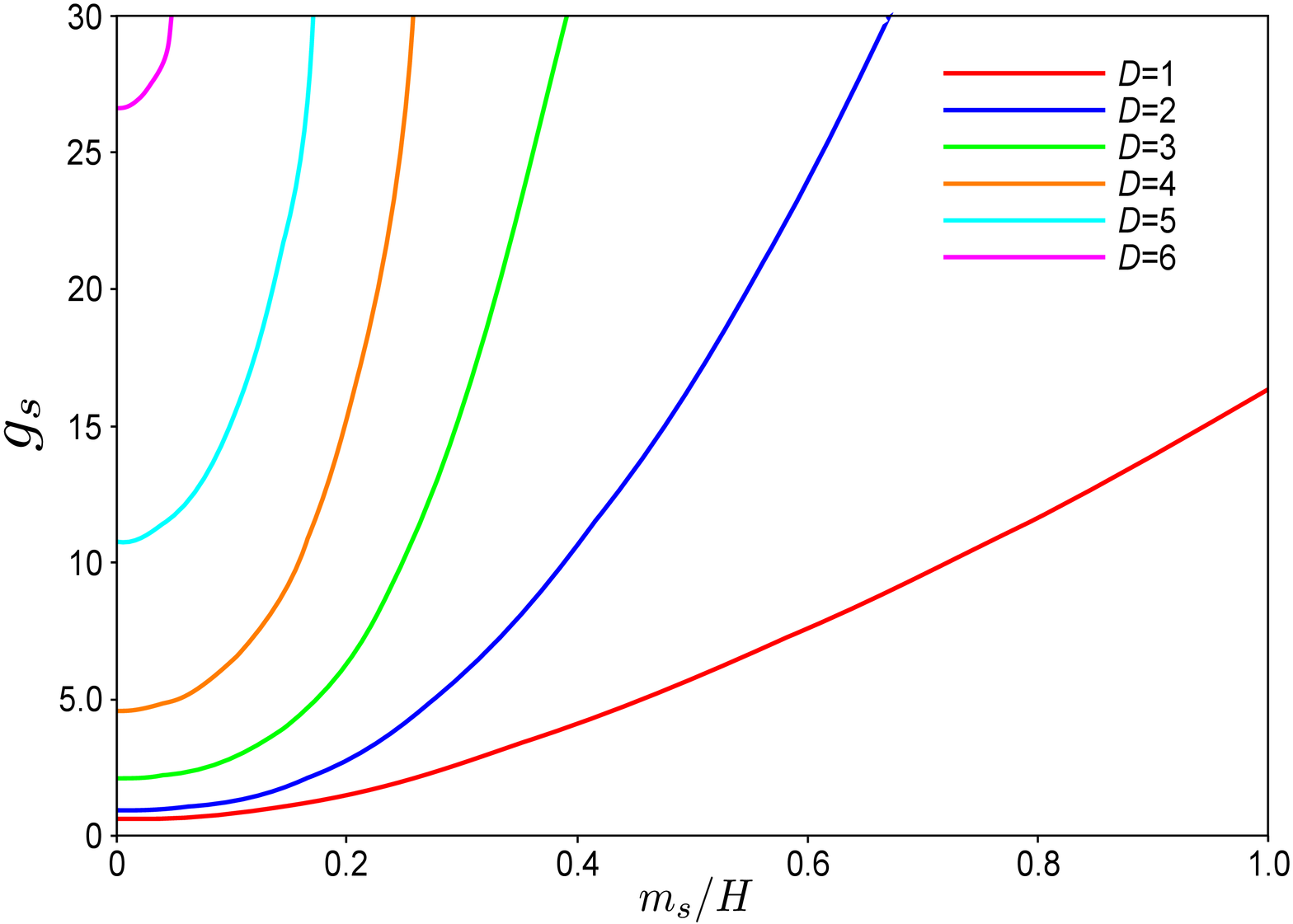}
}
\end{center}
\caption{$g_{s}$ as a function of $m_{s}/H$, with  
$\delta_{T}=2.6\times 10^{-5}$, 
$r_{\rm t/s}=0.22$, $RH\sim 10^{67}$. }
\label{gsms}
\end{figure}

Figs.~\ref{Lms} and \ref{gsms} 
show $(Lm_{pl})$ and $g_{s}$ as functions of $m_{s}/H$, 
respectively. 
It is easier to have weak coupling 
with small $D$,  
while it is easier to keep $L$ not too large 
with large $D$.  
Typical values that are consistent with (\ref{WMAP})
would be: 
\begin{align}
&\{D=2, m_{s}/H=0.2, Lm_{pl}=10^{12}, g_{s}=3\},\\
&\{D=3, m_{s}/H=0.2, Lm_{pl}=10^{10}, g_{s}=5\},\\
&\{D=4, m_{s}/H=0.1, Lm_{pl}=10^{9}, g_{s}=7\}. 
\end{align}
The number of the fields that participate in $\delta T/T$ is
roughly $N\sim (Lm_{s})^{D}$. For the above choice of parameters,
$10^{14}\lesssim N\lesssim 10^{16}$.

\section{Time-dependent Hubble}
So far we have studied the fluctuations generated during inflation,
assuming the background is pure de Sitter. Let us now consider the
effect of time-dependent Hubble. 

\subsection{Spectral index}
\begin{figure}[htb]
\begin{center}
\rotatebox{-90}{
\includegraphics[height=10cm]{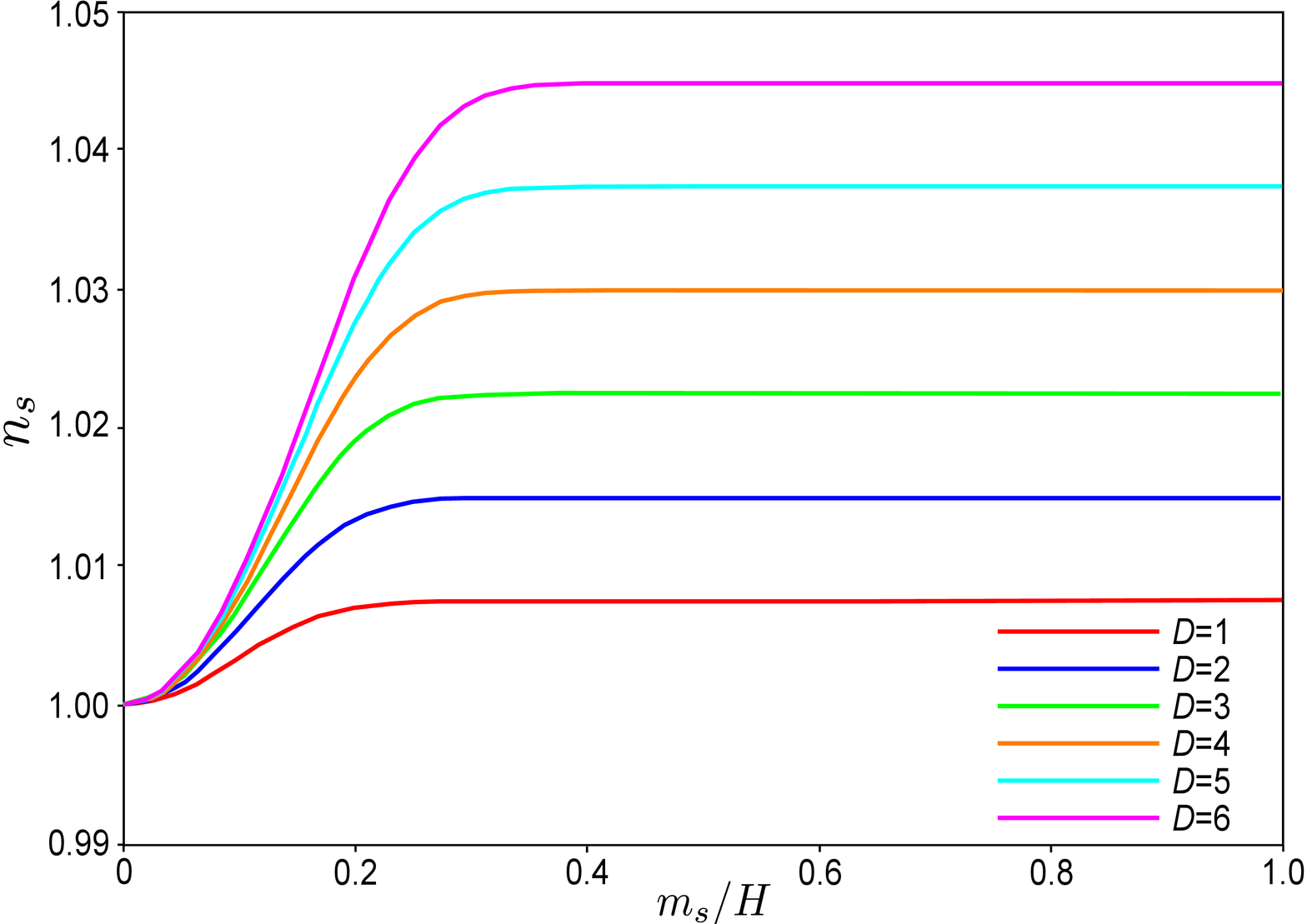}
}
\end{center}
\caption{Spectral index as a function of $m_s/H$.}
\label{index}
\end{figure}

Since $\delta T/T$ originates from the fluctuations of
massive fields, the spectrum is stronger in the
UV. The spectral index $n_s$ is slightly larger than 1, 
\begin{eqnarray}
 n_{s}&=&1-{d\over d\log(Ha|\vec{x}-\vec{x'}|)}\log
\Big\langle{\delta T\over T}(\tau, \vec{x})
{\delta T\over T}(\tau, \vec{x'})\Big\rangle\nonumber\\
&=&1+{4m_{s}^{2}\over 3H^{2}}
{{\cal M}_{D}(\zeta_{0})\over {\cal M}_{D-2}(\zeta_{0})},
\end{eqnarray}
This is in the range $1\lesssim n_{s}\lesssim 1.02$
when $D=2$, and $1\lesssim n_{s}\lesssim 1.05$ for $D\le 6$
(See Figure~\ref{index}). 

The above values are obtained by assuming the Hubble is constant.
However, $n_{s}$ is sensitive to the time-dependence of $H$. 
As long as the change of Hubble is adiabatic $|\dot{H}|/H^2\ll 1$,
it would be reasonable to assume the amplitudes are determined
in terms of $H$ at the time of horizon crossing. We will have to
replace the prefactor $(H/m_{pl})^4$ in the $\langle\Phi\Phi\rangle$
two-point function (\ref{integral}) to $(H(t_{\rm hor})/m_{pl})^4$ where
$t_{\rm hor}$ is the time of the horizon exit for the scale of 
interest ($e^{-Ht_{\rm hor}}\sim |x-x'|$). 
Since the long wave length mode exits the horizon early,
the amplitude is lifted in the infrared. The spectral index
is lowered by $0.5$ if there is time-dependence of order
$\dot{H}/H^2\sim -0.01$.

It would be necessary to understand the origin of vacuum
energy during inflation to understand its time-dependence.
In this paper, we cannot make a definitive statement, but
we would like to mention a possible origin of vacuum energy
in the next subsection.

\subsection{Possible origin of vacuum energy}

In most inflationary models, the presence of vacuum energy
(or a nearly flat inflaton potential) is simply assumed
and its origin is not clear. Also, in the recent constructions
of de Sitter vacua in string theory, the mechanism for 
uplifting from supersymmetric vacua to de Sitter vacua 
is not fully understood. In the study of low energy 
effective action of string theory, it has been very 
difficult to find de Sitter vacua in a controllable 
approximation. (See e.g. \cite{TyeSumitomo} for a 
recent discussion.)

It might be necessary to understand vacuum fluctuations 
of the fields to find de Sitter vacua. In string
compactification models with large internal space~\cite{BBCQ},
which is believed to be realizable generically, there
are many light KK fields. Quantum fluctuations of these
fields might be an important source of vacuum energy. 


With this motivation in mind, in this subsection, we discuss 
a possible dynamical scenario in which vacuum fluctuations and
Hubble are determined in a self-consistent manner.



Consider the expectation value (one-point function) of 
the energy-momentum
tensor. This quantity is UV divergent, and we will renormalize
it so that it  vanishes in the flat background. 
Because of de Sitter
symmetry, the expectation value is proportional to $g_{\mu\nu}$.

The renormalized expectation value of energy-momentum 
tensor of a scalar field in de Sitter background is given by 
(see (6.183) of \cite{BirrellDavies})
\begin{align}
 \langle T_{\mu\nu}\rangle_{\rm ren}
&={g_{\mu\nu}\over 64\pi^2}\Big[ m^2\left\{
m^2+(\xi-{1\over 6})R\right\}\left\{
\psi({3\over 2}+\nu)+\psi({3\over 2}-\nu)
-\log(12 m^2 R^{-1})\right\}\nonumber\\
&\quad -m^2(\xi-{1\over 6})R
-{1\over 18}m^2 R -{1\over 2}(\xi-{1\over 6})^2 R^2
+{1\over 2160}R^2\Big]
\end{align}
where $\psi(z)=\Gamma'(z)/\Gamma(z)$, and $\nu$ is
the order of Hankel function as described in Section 2.  
$\langle T_{\mu\nu}\rangle_{\rm ren}$ is renormalized
by subtracting the divergent piece in the flat background. 
When there are $N$ scalar fields (assuming massless minimally
coupled), we have 
\begin{equation}
 \langle T_{\mu\nu}\rangle_{\rm ren}=N{61\over 960\pi^2}H^{4}g_{\mu\nu}.
\end{equation}
It would be possible that de Sitter space during inflation is
a self-consistent solution of 
\begin{equation}
 R_{\mu\nu}-{1\over 2}g_{\mu\nu}R=-8\pi G\langle T_{\mu\nu}\rangle_{\rm ren}.
\label{selfconsistent}
\end{equation}
The condition that both sides of the equation balances implies
\begin{equation}
 H^2\sim N {H^{4}\over m_{pl}^2}, \quad \Rightarrow \quad 
{H\over m_{pl}}\sim {1\over \sqrt{N}}.
\end{equation}

In fact, our scenario is not consistent with this equation as it is.
The value of $N$ which produces the observed level of
temperature fluctuations is at least $N\sim 10^{12}$
with $H/m_{pl}\sim 10^{-4}$, and $N$ is too large by a 
factor of $10^{4}$ for (\ref{selfconsistent})
to be satisfied. However, this is not a contradiction. 
We do not expect that Einstein equation
(\ref{selfconsistent}) is applicable, since our preferred value
of string coupling is $m_{s}/H\sim 0.1$, and the left hand
side will be corrected by string ($\alpha'$) effects, which 
are not negligible when $m_{s}/H\lesssim 1$. 

Presumably, the self-consistent de Sitter solution of 
(\ref{selfconsistent}) is an unstable solution, and small
fluctuation of $H$ will drive the background to flat space,
which is another solution of this equation. 
To study time-dependence, we will have to compute the expectation 
value $\langle T_{\mu\nu}\rangle_{\rm ren}$
in the background with $\dot{H}\neq 0$. This with (\ref{selfconsistent})
will tell us the evolution of Hubble. We will leave this analysis
as an important open question. 

It is not clear whether the dynamics of quantum expectation
value $\langle T_{\mu\nu}\rangle_{\rm ren}$ is similar to
the dynamics of inflaton, but let us assume it is for the moment. 
In Appendix~A, we perform 
the analysis of fluctuations including inflaton fluctuations 
$\delta\varphi$. 
We take $\delta \varphi$ to be of the same order as $\Phi$, $\Psi$. 
The gravitational potential $\Phi$
has a term induced by $\delta\varphi$ in addition to the term 
from the matter fields that we have studied. 
The relative importance of inflaton and the matter fields
depends on the details,
 such as the slope of the inflaton potential and 
the time between horizon crossing and the end of inflation. 
The effect of the matter fields will be important
unless the slope is fine tuned to a small value.

\section{Non-gaussianities}
Non-gaussianities appear in a characteristic manner in our
mechanism. We first describe the calculation ignoring
interactions among the matter fields $\phi$. 
We then remark that the magnitude of non-gaussianities is 
controlled by the coupling constant in higher dimension. 

The three-point function of $\Phi$ is 
given by the triangle diagram where each pair of 
points is connected by $\langle\phi\phi\rangle$ (see Figure~7),
\begin{eqnarray}
&& \langle \Phi(\tau, \vec{x})\Phi(\tau, \vec{y})\Phi(\tau, \vec{z})\rangle
\label{threepoint}\\
&&\quad ={1\over 8\pi^{3}}\left({H\over m_{pl}}\right)^{6}\sum
\left(H^{3}a^{3}|\vec{x}-\vec{y}|
|\vec{y}-\vec{z}||\vec{x}-\vec{z}|\right)^{-\gamma}.\nonumber
\end{eqnarray}
We define the non-linearity parameter $f_{\rm NL}$ by a local 
replacement, $\Phi\to \Phi_{g}+f_{\rm NL}\Phi_{g}^2$~\cite{Komatsu},
with a gaussian field $\Phi_{g}$. 
Let us consider three-points at superhorizon separation,
and estimate $f_{\rm NL}$ by expansing the coordinates
as in (\ref{expansion}). 
The local form of $f_{\rm NL}$ is enough to characterize the 
magnitude of non-gaussianity in this approximation. 
From (\ref{threepoint}),  
\begin{equation}
 f_{\rm NL}\sim {1\over 24}r_{\rm t/s}\left({m_{s}\over H}\right)^{2}
{{\cal M}_{D}({3\over 2}\zeta_{0})\over 
{\cal M}_{D-2}(\zeta_{0})}.
\end{equation}

\begin{figure}[htb]
\begin{center}
\rotatebox{-90}{
\includegraphics[height=10cm]{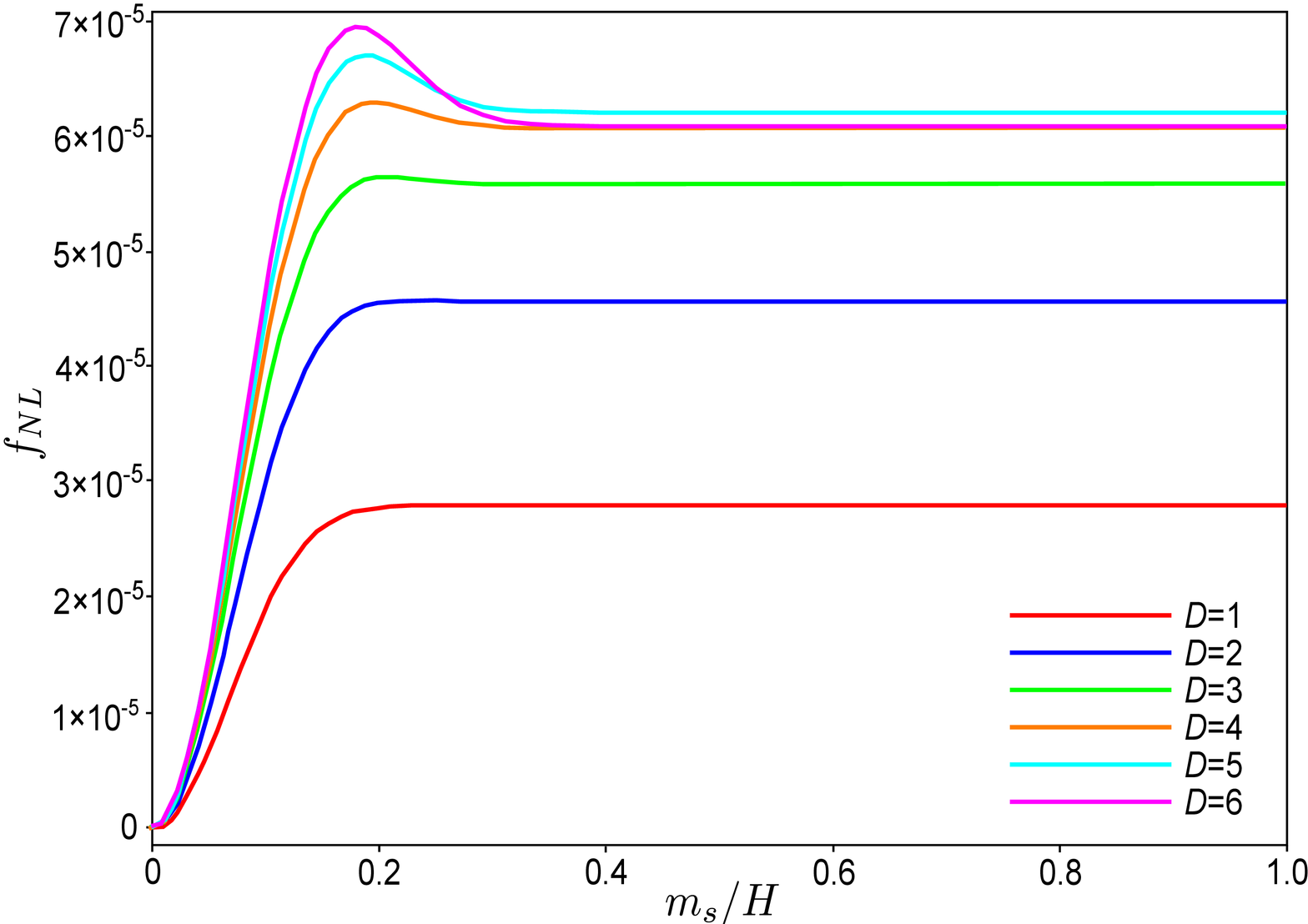}
}
\end{center}
\caption{Non-gaussianity $f_{\rm NL}$ as a function of $m_s/H$.}
\label{nongauss}
\end{figure}

This is proportional to $r_{\rm t/s}$, and further suppressed
by the other factors (See Fig~\ref{nongauss}).
For $r_{\rm t/s}=0.22$, we have $f_{\rm NL}< 10^{-4}.$ 

The reason for the smallness of non-gaussianity is that 
$\langle\Phi\Phi\rangle$ is roughly proportional to the 
number of fields $N$, and $\langle\Phi\Phi\Phi\rangle$
is also proportional to $N$ in our setting. This makes
the non-gaussianity small 
$f_{\rm NL}\sim \langle\Phi\Phi\Phi\rangle/\langle\Phi\Phi\rangle^2
\sim N^{-1}$ in the large $N$ limit. This is in contrast
to the curvaton case~\cite{curvaton}, where 
non-gaussianity is necessarily large if
curvaton is the only source of curvature fluctuations~\cite{curvaton2}. 

\begin{figure}[htb]
\begin{center}
\includegraphics[height=5cm]{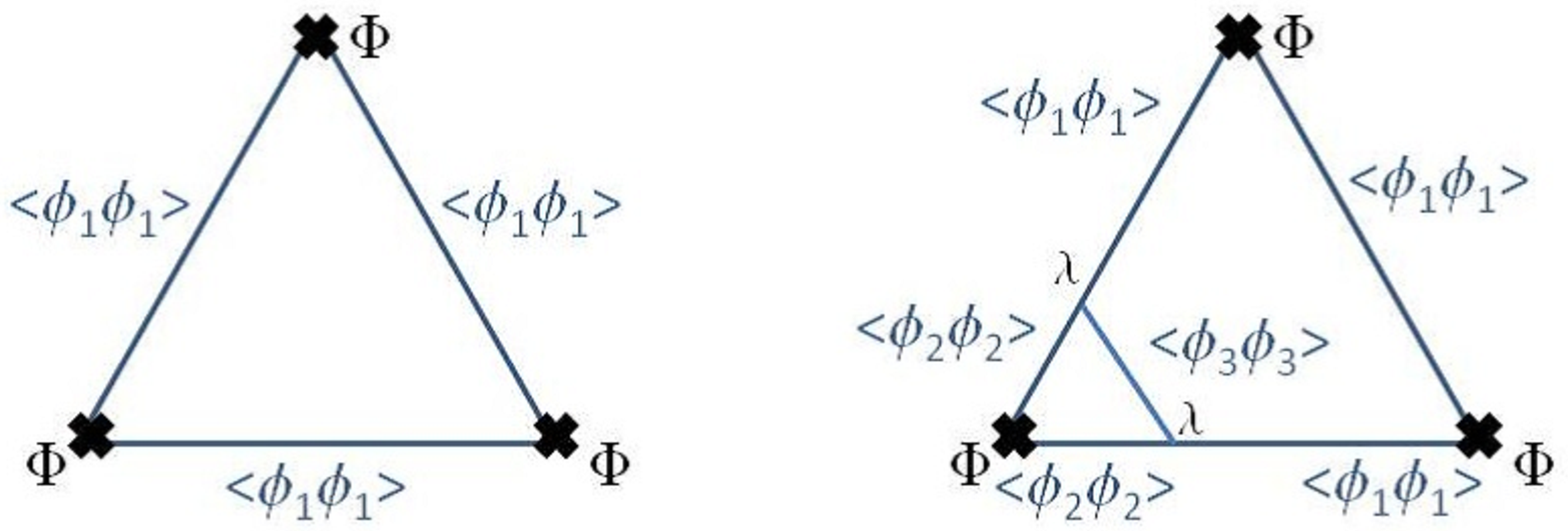}
\end{center}
\caption{Diagrams for the three-point function of $\Phi$. 
Left panel: Ignoring the interaction among the fields $\phi$, 
this is of order $N$ (the number of fields). 
Right panel: Higher loop diagram, where
$\phi_1$ and $\phi_2$ are independent, but $\phi_3$ will
be determined from the conservation law (such as KK momentum
conservation). This is of order $\lambda^2 N^2$.}
\label{triangle}
\end{figure}
When there are interactions among $\phi$'s, we will have
higher loop diagrams, such as the ones in which a propagator 
traverses two sides of the triangle. (See Figure~7.)
Even though there are many fields, interactions do not 
necessarily make $f_{\rm NL}$ huge. Since the 
fields $\phi$ are KK modes or string states, there will be 
conserved quantities such as momentum in the internal space 
or the excitation number of strings. The third field in the
right diagram in Figure~7 is determined by the first two,
and this diagram will be of order 
$\lambda^2 N^2$. The natural magnitude of the coupling
$\lambda$ will be $\lambda_0/\sqrt{V}$ where $\lambda_0$
is the coupling constant in higher dimension, and $V$ is
the volume of internal space. This is because the canonically
normalized KK fields $\phi$ has a factor $1/\sqrt{V}$
relative to the higher dimensional field, and the 
interaction $\lambda \phi^3$ has one more $\phi$ than
the kinetic term. Since $V\sim N$ as mentioned in Section
4.1, the factor $\lambda^2 N^2$ associated to the right diagram
of Figure~7 is just given by $\lambda_0^2$. 
(This is equivalent to saying that the correlation functions
are computed by Feynman rules in the higher dimensions with the 
diagrams of the type of Figure~7.)

If $\lambda_0$ takes finite but small enough value that perturbation
theory is applicable, non-gaussianities will be given by
the diagrams such as the ones in Figure~7. The shape 
(momentum dependence)~\cite{shape} of this type of non-gaussianity will
be different from those arising from the usual slow-roll 
inflation~\cite{Maldacena}. 
The difference will be seen by studying the three-point
correlations at subhorizon separations. We will leave this
analysis to future study.

\section{Conclusions}

We have shown that the CMB temperature fluctuations (adiabatic
perturbation) can be generated by the purely quantum effects of 
fields which are classically at rest. 
When there are a large number of fields, this can produce observable 
level of fluctuations. Tensor fluctuations are hardly affect by
this effect, and will remain of order $H/m_{pl}$. 
In our mechanism, the enhancement of scalar fluctuations relative 
to tensor fluctuation is due to the large number of fields 
involved, and not due to
the smallness of the slow-roll parameter as in the
usual slow-roll scenario.

When the size of the extra dimensions are large 
compared to the inverse Hubble during inflation, we have a large
number of Kaluza-Klein 
modes which contribute to this effect. 
String excited modes also contribute
if $m_s< H$. We compare our results with observed amplitude,
and find that $m_s/H\sim 0.1$ is preferred. The size of extra
dimensions is typically of order $10^7$ GeV$^{-1}$ or smaller.

There have been models of inflation based on TeV scale supersymmetry 
(see \cite{Mazumdar} for a review). Inflation and reheating have
been studied in explicit string compactification in \cite{Cicoli}:
Supersymmetry is broken by the
hidden-sector branes wrapped around internal cycles, and
inflaton is given by closed string moduli. In that case,
it is important to make sure that the decay of inflaton
reheats the visible sector dominantly and not
the hidden sectors significantly, which has been checked
in \cite{Cicoli}. This is necessary to avoid cosmological 
problems, such as the generation of large isocurvature perturbations. 

In our case, we do not assume low-energy supersymmetry. 
In this case, it will be generically the case that 
massive fields, such as KK modes, decay and reach thermal
equilibrium. This will occur well before the standard big bang 
nucleosynthesis begins. In this situation, there will be no 
isocurvature perturbations, as shown in \cite{Weinberg}.

We performed our analysis assuming the extra dimensions are
compactified on a torus $T^D$. We believe this captures
the qualitative features of quantum effects of 
KK states in the general compactifications with $L\gg H^{-1}$,
whenever the multiplicity of the KK modes is similar to that
for $T^D$. There have been studies on string compactifications 
which realizes supersymmetric vacua with all moduli fixed. 
It is argued that the large-volume compactification is 
generically achievable in the construction of \cite{BBCQ}.   
Understanding of the mechanism for uplifting to de Sitter vacua 
or realizing inflation is at a more qualitative level at present.  
Brane anti-brane pair~\cite{Tye} will be a candidate for
such a mechanism. We expect our mechanism for generating
CMB temperature fluctuations should be relevant in these
contexts.

Non-gaussianity in our mechanism is given by triangle diagrams,
with possible corrections. The magnitude
is controlled by the coupling constant in higher dimension. 
It would be possible in principle to distinguish our mechanism 
from others by the precise measurement of non-gaussianities.

The main purpose of this paper is to study fluctuations without
asking the origin of vacuum energy during inflation.
But as we mentioned in Section~6, it would be possible that vacuum 
fluctuations (the renormalized expectation value of energy-momentum 
tensor) of a large number of fields are the source of vacuum energy. 
It is important to understand the dynamics of this 
vacuum energy~\cite{WorkinProgress}.
We have included the analysis of fluctuations in the background 
with time-dependent Hubble by introducing inflaton in the appendix,
but it is not clear to what extent this captures the dynamics of 
quantum vacuum energy.

\subsection*{Acknowledgements}
We thank Misao Sasaki for helpful comments.
The work of H.K is supported by the MEXT Grant-in-Aid for the Global COE 
Program, ``The Next Generation of Physics Spun from Universality and 
Emergence,'' and the JSPS Grant-in-Aid for Scientific Research No. 22540277.
M.N is supported by the JSPS Grant-in-Aid for Scientific Research Nos.
21540290 and 23540332. 
Y.S is supported by JSPS Grant-in-Aid for Scientific Research (A)
No. 23244057, and the MEXT
Grant-in-Aid for Young Scientists (B) No.~21740216. 
M.N acknowledges the Niels Bohr Institute for their hospitality extended
during his stay.

\section*{Appendix}

\appendix

\section{Perturbations including inflaton}
In this appendix, we assume the dynamics of time-dependent Hubble
is effectively described by an inflaton field $\varphi$, and 
study fluctuations including the inflaton fluctuation $\delta\varphi$
in addition to matter fields $\phi$.

We assume inflaton and matter are not directly coupled with
each other. Thus, the equation of motion for $\phi$
is the same as before, except that Hubble is now time-dependent
and ${\cal H}$ and $a$ are not those for pure de Sitter,
\begin{equation}
 \left[\partial_{\tau}^{2}+2{\cal H}\partial_{\tau}
-\bigtriangleup +a^{2}m^{2}\right]\phi=0.
\end{equation}
This does not couple to fluctuations of other fields, so the
quantization of $\phi$ can be done at first. 

The action of inflaton is
\begin{equation}
 S=\int d^{4}x \sqrt{-g}\left({1\over 2}g^{\mu\nu}\partial_{\mu}\varphi
\partial_{\nu}\varphi-V(\varphi)\right).
\end{equation}
We decompose $\varphi$ into classical part (which is homogeneous in
space) and fluctuations,
\begin{equation}
 \varphi=\varphi_{0}(\tau)+\delta\varphi(\tau, \vec{x}).
\end{equation}
Classical part satisfies the equation of motion
\begin{equation}
 \partial_{\tau}^{2}\varphi_{0}+2{\cal H}\partial_{\tau}\varphi_{0}
+a^2V_{,\varphi}=0.
\end{equation}
To find the equation of motion for $\delta\varphi$, it is
convenient to take the longitudinal gauge,
\begin{equation}
 ds^2=a^2\left[(1+2\Phi)d\tau^2-(1-2\Psi)d\vec{x}^2\right],
\end{equation}
and expand the equation of motion,
\begin{equation}
{1\over \sqrt{-g}}\partial_{\mu}\left(\sqrt{-g}g^{\mu\nu}\partial_{\nu}
\varphi\right)+V_{,\varphi}=0,
\end{equation}
to the first order in $\delta\varphi$, $\Phi$, $\Psi$. Then, we get
\begin{equation}
 \left[\partial_{\tau}^{2}+2{\cal H}\partial_{\tau}-\bigtriangleup
+a^{2}V_{,\varphi\varphi}\right]\delta\varphi
-(\varphi'_{0})(\Phi'+3\Psi')
+2a^{2}V_{,\varphi}\Phi=0.
\label{inflatoneom}
\end{equation}

Energy-momentum tensor for $\varphi$ is 
\begin{equation}
 T^{(\varphi) \mu}{}_{\nu}=\partial^{\mu}\varphi\partial_{\nu}\varphi
-{1\over 2}(\partial^{\rho}\varphi\partial_{\rho}\varphi-2V(\varphi))
\delta^{\mu}_{\nu}.
\end{equation}
The classical part (zeroth order in $\delta\varphi$) is
\begin{eqnarray}
 T^{(\varphi, 0)0}{}_{0}&=&{1\over 2a^2}\left((\varphi'_{0})^{2}
+2V\right),\\
 T^{(\varphi, 0)i}{}_{j}&=&{1\over 2a^2}\left(-(\varphi'_{0})^{2}
+2V\right)\delta^{i}_{j}.
\end{eqnarray}
The linear part in $\delta\varphi$ is 
\begin{eqnarray}
\delta T^{(\varphi)0}{}_{0}&=&{1\over a^2}
\left(-(\varphi'_{0})^{2}\Phi+\varphi'_{0}\delta\varphi'
+a^2V_{,\varphi}\delta\varphi\right),\\
\delta T^{(\varphi)0}{}_{i}&=&{1\over a^2}\varphi'_{0}
\partial_{i}\delta\varphi,\\
\delta T^{(\varphi)i}{}_{j}&=&{1\over a^2}
\left((\varphi'_{0})^{2}\Phi-\varphi'_{0}\delta\varphi'
+a^2V_{,\varphi}\delta\varphi\right)\delta^{i}_{j}
\end{eqnarray}
Note that $\delta T^{(\varphi)i}{}_{j}$ does not have
off-diagonal components.

To write Einstein equations, we include the linear terms in
$\delta\varphi$ and the quadratic terms in $\phi$ in the 
energy-momentum tensor. 
The (0,0), (0, $i$) and ($i$, $j$) components of
Einstein equations are as follows:
\begin{eqnarray}
&&\Lap\Psi-3\CH (\Psi'+\CH\Phi)=
4\pi G\Big\{\varphi_{0}'\delta\varphi'
-(\varphi_{0}')^{2}\Phi +{a^{2}}V_{,\varphi}
\delta\varphi \nonumber\\
&&\hspace{5cm} + \sum {1\over 2}(\phi'{}^2
+\partial_{i}\phi\partial_{i}\phi +m^{2}a^{2}\phi^2)
\Big\}
\label{E00I}\\
&&(\Psi'+\CH\Phi),{}_{i} =4\pi G
\Big\{\varphi_{0}'\delta\varphi 
+\sum {1\over \bigtriangleup} 
\partial_{k}(\phi'\partial_{k}\phi)\Big\}_{, i}
\label{E0iI}\\
&&\left[\Psi''+\CH (2\Psi+\Phi)'+(2\CH'+\CH^2)\Phi
+{\Lap\over 2}(\Phi-\Psi)\right]\delta_{ij}
-{1\over 2}(\Phi-\Psi),{}_{ij}\nonumber\\
&&\qquad =4\pi G\Bigg[\left\{(\varphi_{0}')^{2}\Phi
-\varphi_{0}'\delta\varphi'+{a^{2}}V_{,\varphi}
\delta\varphi
+\sum {1\over 2}(\phi'{}^2-\partial_{i}\phi\partial_{i}\phi
-m^2 a^{2}\phi^2)\right\}\delta_{ij}\nonumber\\
&&\qquad \qquad +\sum \left\{{3\over 2\bigtriangleup^{2}}
\partial_{k}\partial_{l}(\partial_{k}\phi\partial_{l}\phi)
-{1\over 2\bigtriangleup}\partial_{k}\phi\partial_{k}\phi
\right\}_{,ij}\Bigg]
\label{EijI}
\end{eqnarray}
where the summation is taken over the species of matter fields
$\phi$, as in the main text.

Let us study the leading behavior in the super-horizon limit 
following \cite{Mukhanov}. We assume inflaton is
classically slow-rolling,
\begin{equation}
 3H\dot{\varphi}_{0}+V_{,\varphi}=0.
\end{equation}
We consider inflaton e.o.m.\ (\ref{inflatoneom}) and the
(0, $i$) component of Einstein equation (\ref{E0iI}). 
In terms of physical time $t$ (and in the slow-roll limit),
\begin{eqnarray}
&& 3H\dot{\delta\varphi}+V_{,\varphi\varphi}\delta\varphi
+2V_{,\varphi}\Phi=0,\\
&&H\Phi=4\pi G\left\{\dot{\varphi}_{0}\delta\varphi
+\sum {1\over \bigtriangleup}\partial_{k}(\dot{\phi}
\partial_{k}\phi)\right\},
\end{eqnarray}
where we have ignored the terms in the energy
momentum tensor of $\phi$ which are small in the small mass
limit, $\gamma={2\over 3}m^{2}H^{-2}\ll 1$ (we ignore the term
$s'$ in (\ref{E0i2})). 

These equations imply
\begin{equation}
 {d\over dt}\left(\delta\varphi {V\over V_{,\varphi}}\right)
-\sum {\gamma H\over 2}f(\phi)e^{-\gamma Ht}=0,
\label{slowroll1}
\end{equation}
where we have defined $f(\phi)$ by
\begin{equation}
 {1\over \bigtriangleup}\partial_{k}(\dot{\phi}
\partial_{k}\phi)=-{\gamma H\over 2}f(\phi)e^{-\gamma Ht}.
\end{equation}
That is,
\begin{equation}
 f(\phi)e^{-\gamma Ht}\sim {1\over 2}\phi^2
\end{equation}
in the late time limit. From (\ref{slowroll1}),
\begin{equation}
 \delta\varphi= {V_{,\varphi}\over V}
\left(C-{1\over 2}\sum f(\phi) e^{-\gamma Ht}\right)
\end{equation}
with a constant $C$. We fix $C$ so that the amplitude
of $\delta\varphi$ is $H$ at the horizon exit (since
$\delta\varphi$ is essentially a massless scalar
inside the horizon),
\begin{equation}
 \delta\varphi= {V_{,\varphi}\over V}\left\{
H_{*} \left({V\over V_{,\varphi}}\right)_{*}
+{1\over 2}\sum f(\phi)(e^{-\gamma Ht_{*}}
-e^{-\gamma Ht})\right\},
\label{deltavarphi}
\end{equation}
where the star denotes the quantities evaluated at the horizon
exit.

In the usual slow-roll inflation, there is only the first term.
The factor $\left({V_{,\varphi}\over V}\right)$ generally
grows towards the end of inflation, and it is 
assumed to be of order $m_p$ 
at the end of inflation. 
The second term represents the effect of 
matter fields to the evolution of inflaton fluctuation.

In terms of $\Phi$, 
\begin{eqnarray}
 \Phi&=&-2\left({V_{,\varphi}\over V}\right)^{2}
\left\{H_{*} \left({V\over V_{,\varphi}}\right)_{*}
+{1\over 2}\sum f(\phi)(e^{-\gamma Ht_{*}}
-e^{-\gamma Ht})\right\}\nonumber\\
&&-{4\pi G\over H}\sum {\gamma H\over 2}f(\phi)e^{-\gamma Ht}\\
&=&-2\left({V_{,\varphi}\over V}\right)^{2}
\left\{H_{*} \left({V\over V_{,\varphi}}\right)_{*}
+{1\over 4}\sum (\phi^{2}_{*}-\phi^{2})\right\}-\sum \pi G\gamma \phi^{2}
\label{slowrollPhi}
\end{eqnarray}
The first term is the part induced from the inflaton 
fluctuation (\ref{deltavarphi}) through the usual mechanism.
The second term is the effect of matter (agrees with the
formula that we have obtained), which exists
even if there is no inflaton.

This expression will be valid until the end of inflation.
After inflation, $\Phi$ will be constant (assuming 
$\phi$ is classically oscillating, in which case it can
be regarded as matter, or $\phi$ has decayed into radiation).
Relative importance of the effect of $\phi$ compared to that
of inflaton in (\ref{slowrollPhi}) depends on the details 
such as how steep the potential is
or how much time has passed between horizon exit and the end
of inflation.

\end{document}